\documentclass[10pt,journal,twoside]{IEEEtran}
\usepackage{amsmath,epsfig,times}
\usepackage{latexsym,epsf,graphicx,psfrag,amssymb,cite}
\usepackage{cite}
\usepackage{amsmath,tikz}
\usepackage{amssymb}
\usepackage{epsfig}
\usepackage{graphicx}
\usepackage{subfigure}
\usepackage{multirow}
\usepackage{array}
\usepackage{enumerate}
\begin{document}

\title{On the Performance of Selection Cooperation with Outdated CSI and Channel Estimation Errors}
\author{
Mehdi~Seyfi,~\IEEEmembership{Student Member,~IEEE,}
Sami~(Hakam)~Muhaidat*,~\IEEEmembership{Member,~IEEE,}
and Jie~Liang,~\IEEEmembership{Member,~IEEE,}
\thanks{The work of Jie Liang was supported in part by the Natural Sciences and Engineering
Research Council (NSERC) of Canada under grants RGPIN312262-05, EQPEQ330976-2006, and
STPGP350416-07. }
\thanks{M. Seyfi, S. Muhaidat and J. Liang are with the School of Engineering
Science, Simon Fraser University, Burnaby, BC, V5A 1S6, Canada. Phone:
778-782-7376. Fax: 778-782-4951. E-mail: msa119@sfu.ca, muhaidat@ieee.org, jiel@sfu.ca.}

\thanks{$^*$Corresponding author
}}
\maketitle

\markboth{IEEE Transactions on Wireless Communications, SUBMITTED: SEPT . 2010. }{M. Seyfi \lowercase{\textit{et al.}}: {Performance Analysis of Selection Cooperation with Feedback Delay and Channel Estimation Errors}}

\def\arg{{\rm arg}}
\def\Re{{\rm Re}}
\def\Im{{\rm Im}}
\def\bA{{\bf A}}
\def\ba{{\bf a}}
\def\bB{{\bf B}}
\def\bb{{\bf b}}
\def\bC{{\bf C}}
\def\bc{{\bf c}}
\def\bD{{\bf D}}
\def\bd{{\bf d}}
\def\bE{{\bf E}}
\def\be{{\bf e}}
\def\bF{{\bf F}}
\def\bG{{\bf G}}
\def\bg{{\bf g}}
\def\bH{{\bf H}}
\def\bh{{\bf h}}
\def\bI{{\bf I}}
\def\bJ{{\bf J}}
\def\bK{{\bf K}}
\def\bk{{\bf k}}
\def\bM{{\bf M}}
\def\bL{{\bf L}}
\def\bN{{\bf N}}
\def\bn{{\bf n}}
\def\bP{{\bf P}}
\def\bp{{\bf p}}
\def\bQ{{\bf Q}}
\def\bq{{\bf q}}
\def\bR{{\bf R}}
\def\br{{\bf r}}
\def\bS{{\bf S}}
\def\bs{{\bf s}}
\def\bT{{\bf T}}
\def\bt{{\bf t}}
\def\bU{{\bf U}}
\def\bu{{\bf u}}
\def\bV{{\bf V}}
\def\bv{{\bf v}}
\def\bW{{\bf W}}
\def\bw{{\bf w}}
\def\bX{{\bf X}}
\def\bx{{\bf x}}
\def\bY{{\bf Y}}
\def\by{{\bf y}}
\def\bZ{{\bf Z}}
\def\bz{{\bf z}}

\def\rE{{\rm E}}
\def\re{{\rho_{\!e}}}
\def\rf{{\rho_{_{\!f}}}}

\def\cC{{\mathcal{C}}}
\def\cE{{\mathcal{E}}}
\def\cI{{\mathcal{I}}}
\def\cN{{\mathcal{N}}}
\def\cS{{\mathcal{S}}}

\def\cH{{\mathcal{H}}}
\def\cE{{\mathcal{E}}}
\def\cI{{\mathcal{I}}}
\def\cN{{\mathcal{N}}}
\def\cS{{\mathcal{S}}}
\def\cD{{\mathcal{D}}}
\def\cY{{\kappa}}
\def\clam{{\left(\frac{\lambda_{_{md}}}{2\sigma^2P\beta}\right)^{(k+1)}}}
\def\NOP{{\sqrt{{\it {\mathcal{E}_{_{si}}}}|\hat{h}_{_{si}}|^2+{\it N_o}}}}
\def\NOB{{\sqrt{{\mathcal{E}_{_{si}}}|\hat{h}_{_{si}}|^2+{\mathcal{E}_{_{id}}}|\hat{h}_{_{id}}|^2+N_o}}}
\def\no{{\nonumber}}
\def\NOP{{\sqrt{{\it {\mathcal{E}_{_{si}}}}|\hat{h}_{_{si}}|^2+{\it N_0}}}}
\def\NOB{{\sqrt{{\mathcal{E}_{_{si}}}|\hat{h}_{_{si}}|^2+{\mathcal{E}_{_{id}}}|\hat{h}_{_{id}}|^2+N_0}}}
\def\no{{\nonumber}}
\def\smo{{_{\!_{sm\!,\!o}}}}
\def\mdo{{_{\!_{md\!,\!o}}}}
\def\ido{{_{_{id\!,\!o}}}}
\def\jdo{{_{_{jd\!,\!o}}}}
\def\ldo{{_{_{ld\!,\!o}}}}
\def\cm{{\hat{\chi}_{\!_{m\!,\!o}}}}
\def\spa{{\!\!\!\!\!}}
\def\msd{{_{\!_{m^{\!\star}\!d}}}}

\begin{abstract}
 We analyze the effect of feedback delay and channel estimation errors in a decode-and-forward (DF) relay selection scenario. Amongst all the relays that decode the source information correctly, only one relay with the best relay-to-destination ($R \to D$) channel quality is selected. More
specifically, the destination terminal first estimates the relay-destination channel state information (CSI) and then sends the index of the best relay  to the relay terminals via a delayed feedback link.  Due to the time varying nature of the underlying channel model, selection is performed based on the old version of the channel estimate. In this paper, we investigate the performance of the underlying selection scheme in terms of outage probability, average symbol error rate (ASER) and average capacity. Through the derivation of an ASER expression and  asymptotic diversity order analysis, we show  that the presence of feedback delay reduces the asymptotic diversity order to one, while the effect of channel estimation errors reduces it to zero. Finally, simulation results are presented to corroborate the analytical results.

\end{abstract}
\section{Introduction}

\PARstart{I}t has been demonstrated that cooperative diversity provides an effective means of improving spectral and power efficiency of wireless networks as an alternative to MIMO systems  \cite{Laneman2004,Nabar2004}. The main idea behind cooperative diversity is that in a wireless environment, the signal transmitted by the source ($S$) is overheard by other nodes, which can be defined as ``relays'' \cite{Giannakis2005}. The source and its partners can then jointly process and transmit their information, thereby creating a ``virtual antenna array'', although each of them is equipped with only one antenna. It is shown in \cite{Anghel2004,Adve2007} that cooperative diversity networks can achieve a diversity order equal to the number of paths between the source and the destination. However, the need of transmitting the symbols in a time division multiplexing (TDMA) fashion reduces the maximum achievable capacity improvements. Additionally, due to power allocation constraints, using multiple relay cooperation is not economical.
To overcome these problems, relay selection is proposed to alleviate the spectral efficiency reduction caused by multiple relay schemes and to moderate the power allocation constraints\cite{Laneman2004,Bletsas2006,Adve2006}.

    In \cite{Bletsas2006}, Bletsas {\it et al.} proposed a fast selection algorithm relative to the coherence time of the channel. For each relay node, they assume timers that are inversely proportional to either the harmonic mean or a $\max-\min$ function of the back-to-back channels. They study the performance of their selection cooperation scheme in the amplify-and-forward mode (S-AF) in terms of the outage probability.
    The outage probability and the average symbol error rate (ASER) expressions of a relay selection with decode-and-forward (S-DF) scenario   have been studied in \cite{Ikki2009} and \cite{Ikki2010}, respectively.
    In \cite{Beres2008}, Beres and Adve analyze relay selection in a network setting, i.e., multi-source networks and introduced a closed-form formula for the outage probability of a single source single destination network with multiple relays in the DF protocol.
    In \cite{Sadek2006,Sreng2003} the nearest neighbor selection for the DF mode is proposed. In their work, the best relay is simply the spatially nearest relay to the base station.
    In \cite{Adve2007,Adve2006}, the relay whose path introduces the maximum SNR is selected, in the amplify-and-forward (AF) scheme. In \cite{Adve2006}, the improvement of S-AF over the all participate cooperation scheme in the AF mode (AP-AF) is discussed in terms of symbol error rate. The ASER of S-AF and AP-AF is discussed based on an approximation of the cumulative density function of the received SNR. In \cite{Adve2007}, Zhao and Adve study the outage probability of S-AF scheme and analyze the performance improvement due to S-AF. They also show dramatic improvements of S-AF over AP-AF in terms of capacity. In \cite{Poor2009} the authors discussed the outage probability, ASER, ergodic capacity and outage capacity in S-DF protocol and compared their results with selection relaying (SR). In \cite{Ikkic2009}, Ikki and Ahmed derive outage probability, ASER, and capacity expressions for S-AF and illustrate the improvement of S-AF over AP-AF. It is shown that the relay links in S-AF as well as AP-AF introduce the same diversity order of $M$, where $M$ is the number of relays.

    \emph{Related work and contributions}:  The effect of feedback delay on the outage probability of antenna selection in a MIMO communications system is discussed in \cite{Ramya2009}. Relay selection in the setups in\cite{Bletsas2006,Beres2008,Ikki2009,Ikki2010,Sadek2006,Sreng2003}, assume the perfect channel state information (CSI) knowelge. However, in practical scenarios,  the communication links are not known and have to be estimated. Since channel estimation is necessary for the selection procedure, channel estimation error due to imperfection of the estimator is inevitable. The effect of channel estimation error in selection cooperation in the AF mode is studied in \cite{Seyfivtc2010,SeyfiQueens2010}. In \cite{Seyfivtc2010}, we analyze the effect of channel estimation error on the outage probability of selection cooperation in the AF mode. Capacity of selection cooperation with imperfect channel estimation is studied in \cite{SeyfiQueens2010, Behbahani2008}.
    The impact of channel estimation error on the ASER performance of distributed space time block coded (DSTBC) systems is investigated in \cite{Cheng2005}, assuming the amplify-and-forward protocol. Building upon a similar set-up, Gedik and Uysal \cite{Gedik2009} extend the work of \cite{Cheng2005} to a system with $M$ relays. In \cite{Han2009}, the symbol error analysis is investigated for the same scenario as in \cite{Gedik2009}.

    In a practical relay selection scenario, the destination is responsible for estimating the  CSI and performing relay selection. The index of the selected relay is then fed back to all the relays. Due to the time varying nature of the fading channels, which is function of $\emph{Doppler shift}$ of the moving terminals, the CSI corresponding to the selected relay is time varying. In this case, relay selection is performed based on  outdated CSI. Hence,  the selection scheme  may not yield  the best relay.

    Although there have been  research efforts on conventional MIMO antenna selection  with feedback delay and channel estimation errors (see for example \cite{Ramya2009} and references therein), only a few isolated results have been reported in the context of cooperative communications. In \cite{Vicario2009}, Vicario \emph{et al.}  analyze the outage probability and the achievable diversity order of an opportunistic relay selection scenario with feedback delay. In this paper, we investigate the impact of feedback delay and channel estimation errors on the performance of a cooperative diversity scheme with relay selection. To the best of our knowledge, this is the very first paper that investigates relay selection with outdated channel estimates. Assuming imperfect channel estimation  and feedback delay, our contributions  are summarized as follows:
\begin{itemize}
  \item We derive exact ASER and outage probability expressions for relay selection with decode-and-forward (S-DF) relaying in the presence of feedback delay and channel estimation errors.
   \item We derive a lower bound on the average capacity in the presence of feedback delay and channel estimation errors.
  \item We demonstrates that the asymptotic diversity order is reduced to one with feedback delay and  reduced to zero in the presence of channel estimation error.
  \item We present a comprehensive Monte Carlo simulation study to confirm the analytical observations and give insight into system performance.
\end{itemize}

    The rest of the paper is organized as follows: In Sec.~\ref{Sec.Sys Model}, we introduce our system setup and the selection strategy. In this section the channel estimation error as well as the delayed feedback link models are illustrated. In Sec.~ \ref{SEC.OP}, we investigate the outage probability of the system and develop a closed-form expression for outage probability. In Sec.~\ref{SEC.ASER}, we study the ASER performance of the system in detail and derive exact expression for the ASER. In Sec.~\ref{SEC.ADO}, we analyze the asymptotic order of diversity in S-DF. In particular, we investigate the effect of channel estimation errors and  feedback delay on the diversity of the system. In Sec.~\ref{SEC.CAPAcity} we introduce a lower bound for the instantaneous capacity and derive a closed-form analytical expression for the average lower bound capacity.  Simulation results are presented in Sec.~\ref{SEC.Simulation R}, and the paper is concluded in  Sec.~\ref{SEC.CoNC}.

\section{system model}\label{Sec.Sys Model}
\begin{figure}[t]
   \begin{center}
    \begin{tabular}{c}
       \psfig{figure=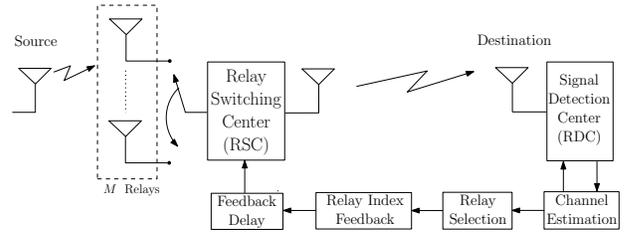,width=3.2in} \\
    \end{tabular}
   \end{center}
 \caption{\spa System model for decode-and-forward relay selection with delayed feedback.}
\label{Fig_M0} \vskip -11pt
\end{figure}

Fig.~\ref{Fig_M0} shows the selection cooperation system model studied in this paper.
We consider a multi-relay scenario with $M$ relays. We assume that the relay $R_m$, $m=1,...,M$, the source $S$, and the destination $D$ are equipped with single transmit and receive antennas, respectively.
In our system model, we ignore the direct transmission
between the source and its destination due to shadowing. $h_{\!_{sm}}$ and $h_{\!_{md}}$ represent the channel fading gains between $S\to R_m$ and  $R_m\to D$, respectively.
Assuming a half duplex constraint, the data transmission is performed in two time slots.
In the first time slot the source terminal transmits its data to all potentially available $M$ relays. After, receiving the source signal via  independent channels, all the relays $R_m$, $m=1, 2, \ldots M$, decode their received signal, and check whether the transmitted signal is decoded correctly or not. This can be done via some ideal  cyclic redundancy codes (CRC) \cite{Merkey1984}, which are added to the transmitted information symbols.
We define the decoding set $\cD(s)$ as the set of relays that decode the transmitted signal correctly.
Clearly only those relay nodes with a good source to relay channel can be in the decoding set $\cD(s)$. In the second time slot, the best relay that satisfies
an index of merit participates in the transmission and broadcasts its decoded symbol towards the destination.

In this paper, conditioned on the decoding set $\cD(s)$, the selection function selects the relay with the best downlink channel i.e.,
\begin{eqnarray}
m^\star&=&\arg~\max_{m\in \cD(s)}\{\gamma_{md}\}\no\\
&=&\arg~\max_{m\in \cD(s)}\{P|h_{_{md}}|^2\},
\end{eqnarray}
where $\gamma_{md}$ is the received SNR from the $m^{th}$ path.
We assume that the source has a power constraint of $P$ Joules/symbol and similarly each relay node in $\cD(s)$ can potentially transmit its information with $P$ Joules/symbol, and the receiver noise power is $1$.

\subsection{model of channel estimation error}\label{SEC.CE}
All  channels are assumed to be independent and identically distributed (i.i.d), with $h\sim\mathcal{CN}(0,\sigma^2_h)$.
A block fading scheme is considered where the channel realizations are assumed to be constant over a block and correlated across blocks. The correlation coefficient between the $k^{th}$ and the $(k+i)^{th}$ block is $\rho_{_f}=J_o(2\pi f_{_d}Ti)$, where $f_{_d}$ is the Doppler frequency, $J_o(\cdot)$ is the zeroth order modified Bessel function of first kind and $T$ is the block duration.
Minimum mean square error (MMSE) estimators are employed at the relay nodes.
We assume that the number of training symbols in each frame is sufficiently less than the data symbols, so that increasing the training symbols' power would not lead to an increase in the average power; hence, we assume that the training symbols' power is independently adjustable.
Let the true channel be $h$ and the estimated channel be $\hat{h}$, then $h$ and $\hat{h}$ are related by\cite{Leung2003}
\begin{subequations}
\begin{eqnarray}
\hat{h}&=&h+e,\\
h&=&\re\hat{h}+u,\label{EQ_estimation_error_model}
\end{eqnarray}
\end{subequations}
where $e$ and $u$ are the channel estimation errors. $\hat{h},~e$ and $u$, are all zero mean Gaussian random variables with variances $\sigma^2_{\hat{h}}$, $\sigma^2_{e}$ and $\sigma^2_{u}$. $\re=(\sigma^2_h/\sigma^{2}_{\hat h})$ is the correlation coefficient between the true channel and the estimated channel. The channel estimation error variances can be written as
\begin{subequations}
\begin{eqnarray}
\sigma^2_u&=&(1-\re)\sigma^2_h,\\\label{EQ_sigma^2}
\sigma^2_e&=&(1-\re)\sigma^2_{\hat{h}}.\label{EQ_sig_e}
\end{eqnarray}
\end{subequations}
\subsection{Delayed feedback model}\label{Sec.DelayedF}
There are two main selection strategies in the literature. In the first scenario, the relay nodes are in charge of monitoring of their individual source and destination instantaneous CSI \cite{Bletsas2006}. In this strategy, CSI is estimated at the relay nodes for further decisions on which relay is the best node to cooperate. In particular, when the relay nodes overhear a ready-to-send (RTS) packet form the source terminal, they estimate the source link CSI, i.e., $h_{\!_{sm\!}}$. Then, upon receiving a clear-to-send (CTS) packet from the destination, the corresponding destination link at the relays, i.e., $h_{\!_{md}}$ is estimated. Right after receiving the CTS packet, the relay nodes ignite a timer which is a function of the instantaneous relay-destination CSI\footnote{The timer is a function of relay-destination CSI for DF mode and source-relay-destination CSI for AF mode cooperation.}.
 The timer of the best relay expires first, and a flag packet is sent to other relays, informing them to stop their timers as the best relay is selected. Then the relays that receive the flag packet, keep silent and the selected relay participates in communication\cite{Bletsas2006}.

In the second strategy, the relay nodes estimate their individual uplink CSI. The relays that correctly decode their received symbols from the source, send a flag packet to the destination, announcing that they are ready to participate in cooperation. The destination terminal, on the other hand, estimates the downlink CSI, orders the received SNR from each relay in the decoding set and feedbacks the index of the best relay that introduces the maximum received SNR via a $\log M$ bit feedback link. The selected relay then operates with full power\cite{Jaffarkhani2009}.

In both strategies, a very important issue that must be taken into consideration is the selection speed. Communication links among  terminals are time varying with a macroscopic rate in the order of \emph{Doppler shift}, which is inversely proportional to the channel \emph{coherence time} \cite{Bletsas2006}. Any relay selection scheme must be performed no slower than the channel coherence time, otherwise selection is performed based on the old CSI, while the channel conditions are altered at the time that selection is performed. This might lead to a wrong selection of relay and therefore affect the performance of the system.

In this paper, to simplify the notation, we adopt the second strategy in relay selection. However, the first strategy obeys the same rules and formulations.

In the second strategy, since the feedback link only transmits the index of the selected relay, a lower feedback bandwidth
is required.

Let the estimated channel be $\hat{h}$ and the old estimated CSI based on which the selection is performed be $\hat{h}_o$. Since $\hat{h}$ and $\hat{h}_o$ are both zero mean and jointly Gaussian they can be related as follows\cite{Ramya2009}

\begin{equation}\label{EQ_hat_est}
\hat{h}=\sigma_{\hat{h}}\left(\frac{\rf}{\sigma_{\hat{h}_o}}\hat{h}_o+\sqrt{1-\rf^2}v\right)
\end{equation}
where $v\sim \cN(0,1)$.
We stress that $\hat{h}_o$ is the channel which is used for relay selection, whereas $\hat{h}$ is the channel which is used for decoding.

\section{Outage Probability of DF relay selection}\label{SEC.OP}

In the first time slot, the source terminal communicates with the relay terminals by broadcasting the signal $x$. Let the received signal at each relay be  $y_{\!_{sm}}$, $m=1,\ldots M$. To decode the received symbol, each relay estimates the corresponding $S\to R_m$ channel and then decodes the received signal using an ML decoder. The old received signal at the $m^{th}$ relay can be written as
\begin{subequations}
\begin{eqnarray}
\hat{x}&=&\sqrt{P}\hat{h}^*{\!\!\!\smo}y_{\!_{sm}}\no\\
       &=&\sqrt{P}\hat{h}^*{\!\!\!\smo}(\sqrt{P}h{\smo}x+n_{\!_{sm}})\label{EQ_h_SM}\\
       &=&\sqrt{P}\hat{h}^*{\!\!\!\smo}\left[\sqrt{P}\left(\re \hat{h}{\smo}+u_{\!_{sm}}\right)x+n_{\!_{sm}}\right]\no\\
       &=&\underbrace{P\re|\hat{h}{\smo}|^2x}_{\textsf{message~component}}\!\!\!\!\!+\underbrace{P\hat{h}^*{\!\!\!\smo}u_{\!_{sm}}x}_{\textsf{error ~component}}+\!\!\!\underbrace{\sqrt{P}\hat{h}^*{\!\!\!\smo}n_{\!_{sm}}}_{\textsf{noise~component}},\label{EQ_Relay_received}
\end{eqnarray}
\end{subequations}
where we have substituted (\ref{EQ_estimation_error_model}) into (\ref{EQ_h_SM}). Using (\ref{EQ_Relay_received}), the received effective SNR at each relay is given by
\begin{equation}\label{EQ_SNR_Relay}
\hat{\gamma}{\smo}^{\rm\!\!\!\!\!\!\! eff}=\frac{P\re^2|\hat{h}{\smo}|^2}{1+P\sigma^2_{u_{\!_{sm}}}}=P\hat{\gamma}{\smo},
\end{equation}
where $\hat{\gamma}{\smo}=\frac{\re^2|\hat{h}{\smo}|^2}{1+P\sigma^2_{u_{\!_{sm}}}}$.
Similarly, in the second time slot, the effective SNR received by the destination via the $m^{th}$ relay  is

\begin{equation}\label{EQ_SNR_destination}
\hat{\gamma}{\mdo}^{\rm\!\!\!\!\!\!\! eff}=\frac{P\re^2|\hat{h}{\mdo}|^2}{1+P\sigma^2_{u_{\!_{md}}}}=P\hat{\gamma}{\mdo},
\end{equation}
where $\hat{\gamma}{\mdo}=\frac{\re^2|\hat{h}{\mdo}|^2}{1+P\sigma^2_{u_{\!_{md}}}}$, given that $m\in \cD(s)$. Here, we assume that $\hat{\gamma}{\smo}$ and $\hat{\gamma}{\mdo}$ are exponentially distributed with parameters $\lambda_{\smo}=(1+P\sigma^2_{u_{\!_{sm}}})/\re$ and $\lambda_{\mdo}=(1+P\sigma^2_{u_{\!_{md}}})/\re$.

If the relay $R_m$ is in the decoding set, by sending a flag packet, it signals its capability of participating in cooperation. Then, based on the old channel realizations, the destination selects the relay with the best $R_m\rightarrow D$ link
\begin{equation}
R_{m^{\!\star}}=\arg \max_{m\in\cD(s)}\{\hat{\gamma}\mdo\}.
\end{equation}

\subsection{Outage probability with feedback delay} Once the best relay is selected by destination, the index of $R_{m^{\!\star}}$ is fed back to all relays via a delayed feedback link. This means, at the time the relays receive the index, the system's CSI  has changed, due to the time varying nature of communication links. Let the current $R_{m^{\!\star}}\rightarrow D$ channel realization, the corresponding estimate and the received SNR from the selected relay be $h_{_{\!m^{\!\star}\!d}}$, $\hat{h}_{_{\!m^{\!\star}\!d}}$ and $\hat{\gamma}_{_{\!m^{\!\star}\!d}}$, respectively.
Due to feedback delay, relay selection is done based on past channel coefficients instead of current coefficients, i.e., $\hat{\gamma}_{_{m^{\!\star}\! d}}=\hat{\gamma}_{_{md}}$ where $m=\arg~\max_{\tiny\!\!\!\!\!\!\!\!\!\!\!\!\!\!\!\!\!\!\!\!\begin{array}{l}\\{i\in\cD(s)}\end{array}}\!\!\!\{\hat{\gamma}_{_{id\!,\!o}}\}$.

Assuming that the communication between source and destination targets an end-to-end data rate $R$, the system is in outage if the $S\rightarrow R_{m^{\!\star}}\rightarrow D$ link observes an instantaneous capacity per bandwidth\footnote{Since the system is exposed to channel estimation error, the instantaneous capacity which is mentioned here, is only a lower bound for the true instantaneous capacity; hence, the derived expression for outage probability is a lower bound for the true probability of outage in the presence of channel estimation error and is exact with the perfect CSI. For further details please refer to Sec. \ref{SEC.CAPAcity}.} $C^{\star}=\frac12\log\left(1+P\hat{\gamma}_{_{\!m^{\!\star}\!d}}\right)$ that is below the required rate $R$, i.e.,
\begin{eqnarray}
P_o\!\!\!&=&\!\!\!Pr\left[\frac12\log\left(1+P\hat{\gamma}_{_{\!m^{\!\star}\!d}}\right)\le R\right]\no\\
\!\!\!&=&\!\!\!\sum_{\tiny \cD(s)}Pr[\cD(s)]\no\\
\!\!\!&\times&\!\!\!Pr\left[\hat{\gamma}_{_{\!m^{\!\star}\!d}}\le R_o,\hat{\gamma}_{_{\!m^{\!\star}\!d\!,\!o}}=\max_{_{\!m\in \cD(s)}
}\{\hat{\gamma}\mdo\}\big|\cD(s)\right],\\\label{EQ_Selection Rule}
&=&\!\!\!\sum_{\tiny \cD(s)}Pr[\cD(s)]\no\\
\!\!\!\!&\!\!\times\!\!&\!\!\!\!\spa\sum_{\tiny m\in \cD(s)}\!\!\!\!\!Pr\left[\hat{\gamma}_{_{md}}\le R_o,\hat{\gamma}_{_{md\!,\!o}}\geq\!\!\!\!\!\!\max_{_{\tiny\begin{array}{l}i\in\!\!\cD(s)\\i\neq m\end{array}}}\!\!\!\!\{\hat{\gamma}_{_{id\!,\!o}}\}\big|\cD(s)\right],\label{EQ_Selection Rule2}
\end{eqnarray}
where $R_o=\frac{2^{2R}-1}{P}$.

\subsection{Probability of decoding set $\cD(s)$} The relay, $R_m$, is in the decoding set $\cD(s)$ if the $S\rightarrow R_m$ link observes an instantaneous capacity per bandwidth $C\smo$ that is above the required rate $R$
\begin{equation}\label{EQ_I_sm}
C{\smo}=\frac12\log_2\left(1+P\hat{\gamma}{\smo}\right)\geq R.
\end{equation}
 Noting that $\hat{\gamma}\smo$ is exponentially distributed, relay $R_m$ is in the decoding set if\cite{Beres2008}
\begin{eqnarray}
Pr\left[R_m\in \cD(s)\right]&=&Pr\left[\hat{\gamma}\smo\geq R_o\right].\no\\
&=&\exp\left(-\lambda\smo R_o\right).\label{EQ_pr_DS}
\end{eqnarray}
Finally, the probability of selecting a specific decoding set is\cite{Beres2008}
\begin{eqnarray}\label{EQ_DS}
Pr[\cD(s)]\!\!\!\!&=&\!\!\!\!\prod_{\tiny m\in\cD(s)}\exp\left(-\lambda\smo R_o\right)\no\\
\!\!\!\!&\times&\!\!\!\! \prod_{\tiny m\notin\cD(s)}\left[1-\exp\left(-\lambda\smo R_o\right)\right].
\end{eqnarray}

\newcounter{c1}
\begin{figure*} [t]
\normalsize \setcounter{c1}{\value{equation}}
\setcounter{equation}{17}
\begin{eqnarray}\label{EQ_Pr(I_SM<R)}
P_o^{\cD(s)}=\sum_{\tiny k=0}^\infty \frac{\lambda\mdo^{\!\!\spa k+1}\gamma(k+1,\lambda_{_{md}}R_o/2\sigma^2)}{(k!)^2}\left(\frac{c_{_m}}{2}\right)^{k}\int_0^{\infty}\hat{\gamma}_{_{md\!,\!o}}^k \left[e^{-(\lambda\mdo+\lambda\mdo\frac{c_{_m}}{2})\hat{\gamma}_{_{md\!,\!o}}}-\spa\sum_{\tiny \tiny\begin{array}{l}i\!\in\!\!\cD(s)\end{array}}\spa e^{-(\lambda\mdo+\lambda\ido+\lambda\mdo\frac{c_{_m}}{2}) \hat{\gamma}_{_{md\!,\!o}}}\right.\no\\+\spa\sum_{\tiny \tiny\begin{array}{l}i,j\!\in\!\!\cD(s)\\i\neq j\end{array}}\spa e^{-(\lambda\mdo+\lambda\ido+\lambda\jdo+\lambda\mdo\frac{c_{_m}}{2}) \hat{\gamma}_{_{md\!,\!o}}}
-\ldots+\left.(-1)^{|\cD(s)|}\spa\spa\sum_{\tiny \tiny\begin{array}{l}i,j,\ldots,l\!\in\!\!\cD(s)\\i\neq j,\ldots, l\end{array}}\spa\spa e^{-(\lambda\mdo+\lambda\ido+\lambda\jdo+\ldots \lambda\ldo+\lambda\mdo\frac{c_{_m}}{2}) \hat{\gamma}_{_{md\!,\!o}}}\right]d_{_{\!\hat{\gamma}_{_{md\!,\!o}}}}.\no
\end{eqnarray}
\begin{eqnarray}
P_o^{\cD(s)}=\sum_{\tiny k=0}^\infty \frac{\gamma(k+1,\lambda_{_{md}}R_o/2\sigma^2)}{(k!)} \left(\frac{c_{_m}}{2}\right)^{k}\left[\frac{\lambda\mdo^{\!\!\spa k+1}}{(\lambda\mdo+\lambda\mdo\frac{c_{_m}}{2})^{k+1}}-\spa\sum_{\tiny \tiny\begin{array}{l}i\!\in\!\!\cD(s)\end{array}}\spa\frac{\lambda\mdo^{\!\!\spa k+1}}{(\lambda\mdo+\lambda\ido
+\frac{c_{_m}}{2})^{k+1}}\right.\no\\
+\spa\sum_{\tiny \tiny\begin{array}{l}i,j\!\in\!\!\cD(s)\\i\neq j\end{array}}\spa\spa \frac{\lambda\mdo^{\!\!\spa k+1}}{~~(\lambda\mdo+\lambda\ido+\lambda\jdo+\lambda\mdo\frac{c_{_m}}{2})^{k+1} }-\ldots
\left.+(-1)^{|\cD(s)|}\spa\spa\sum_{\tiny \tiny\begin{array}{l}i,j,\ldots,l\!\in\!\!\cD(s)\\i\neq j,\ldots, l\end{array}}\spa\spa \frac{\lambda\mdo^{\!\!\spa k+1}}{~~(\lambda\mdo+\lambda\ido+\lambda\jdo+\ldots \lambda\ldo+\lambda\mdo\frac{c_{_m}}{2})^{k+1}} \right].\label{EQ_PO_DS_Final}
\end{eqnarray}
\hrulefill
\begin{eqnarray}
P_o=\sum_{\tiny \tiny\begin{array}{c}\cD(s)\end{array}}\spa\prod_{\tiny \tiny\begin{array}{c}\!m\!\!\in\!\!\!\cD(s)\end{array}}\spa\exp\left(-\lambda\smo R_o\right)
\spa\prod_{\tiny \tiny\begin{array}{c}\!\!m\!\!\notin\!\!\!\cD(s)\end{array}}\spa\left[1-\exp\left(-\lambda\smo R_o\right)\right]\spa\sum_{\tiny \tiny\begin{array}{l}m\!\!\in\!\!\cD(s)\end{array}}\sum_{\tiny k=0}^\infty \frac{\gamma(k+1,\lambda_{_{md}}R_o/2\sigma^2)}{(k!)} \left(\frac{c_{_m}}{2}\right)^{k}\no\\\times\left[\frac{\lambda\mdo^{\!\!\spa k+1}}{(\lambda\mdo+\lambda\mdo\frac{c_{_m}}{2})^{k+1}}-\spa\sum_{\tiny \tiny\begin{array}{l}i\!\in\!\!\cD(s)\end{array}}\spa\frac{\lambda\mdo^{\!\!\spa k+1}}{(\lambda\mdo+\lambda\ido
+\frac{c_{_m}}{2})^{k+1}}\right.
+\spa\sum_{\tiny \tiny\begin{array}{l}i,j\!\in\!\!\cD(s)\\i\neq j\end{array}}\spa\spa \frac{\lambda\mdo^{\!\!\spa k+1}}{~~(\lambda\mdo+\lambda\ido+\lambda\jdo+\lambda\mdo\frac{c_{_m}}{2})^{k+1}}-\ldots\no\\
\left.+(-1)^{|\cD(s)|}\spa\spa\sum_{\tiny \tiny\begin{array}{l}i,j,\ldots,l\!\in\!\!\cD(s)\\i\neq j,\ldots, l\end{array}}\spa\spa \frac{\lambda\mdo^{\!\!\spa k+1}}{~~(\lambda\mdo+\lambda\ido+\lambda\jdo+\ldots \lambda\ldo+\lambda\mdo\frac{c_{_m}}{2})^{k+1}} \right].\label{EQ_OUT_FINAL}
\end{eqnarray}
\hrulefill
\begin{eqnarray}\label{EQ_OUT_FINAL_simp}
P_o=\sum_{l=1}^M\exp\left(-l\lambda R_o\right)\left[1-\exp\left(-\lambda R_o\right)\right]^{M-l}\times l\times\sum_{\tiny k=0}^\infty \frac{\gamma(k+1,\lambda R_o/2\sigma^2)}{(k!)} \left(\frac{c}{2}\right)^{k}\times\sum_{m=1}^l\frac{\left(\!\!\!\begin{array}{c}l-1\\m-1\end{array}\!\!\!\right)}{(m+\frac{c}{2})^{k+1}}.
\end{eqnarray}
\setcounter{equation}{\value{c1}}\hrulefill
\end{figure*}

\newcounter{c2}
\begin{figure*} [t]
\normalsize \setcounter{c2}{\value{equation}}
\setcounter{equation}{25}
\begin{eqnarray}\label{EQ_f_cond}
f_{\hat{\gamma}_{_{\!m^{\!\star}\!d}|\cD(s)}}(x)=\sum_{\tiny k=0}^\infty \frac{\lambda_{_{md}} e^{(-\frac{\lambda_{_{md}}x}{2\sigma^2})}(\frac{\lambda_{_{md}}x}{2\sigma^2})^{k}}{2\sigma^2k!} \left(\frac{c_{_m}}{2}\right)^{k}\times\left[\frac{\lambda\mdo^{\!\!\spa k+1}}{(\lambda\mdo+\lambda\mdo\frac{c_{_m}}{2})^{k+1}}-\spa\sum_{\tiny \tiny\begin{array}{l}i\!\in\!\!\cD(s)\end{array}}\spa\frac{\lambda\mdo^{\!\!\spa k+1}}{(\lambda\mdo+\lambda\ido
+\lambda\mdo\frac{c_{_m}}{2})^{k+1}}\right.\no\\+\left.\spa\spa\sum_{\tiny \tiny\begin{array}{l}i,j\!\in\!\!\cD(s)\\i\neq j\end{array}}\spa\spa \frac{\lambda\mdo^{\!\!\spa k+1}}{~~(\lambda\mdo+\lambda\ido+\lambda\jdo+\lambda\mdo\frac{c_{_m}}{2})^{k+1} }-\ldots\right.
\left.+(-1)^{|\cD(s)|}\spa\spa\sum_{\tiny \tiny\begin{array}{l}i,j,\ldots,l\!\in\!\!\cD(s)\\i\neq j,\ldots, l\end{array}}\spa\spa \frac{\lambda\mdo^{\!\!\spa k+1}}{~~(\lambda\mdo+\lambda\ido+\lambda\jdo+\ldots \lambda\ldo+\lambda\mdo\frac{c_{_m}}{2})^{k+1}} \right].
\end{eqnarray}
\hrulefill
\begin{eqnarray}\label{EQ_FINAL_SER}
\bar{P}_e=\frac1{2}\prod_{\tiny i=1}^MB_i+\spa\spa\sum_{\tiny \tiny\begin{array}{c}\cD(s)\\|\cD(s)|\geq1\end{array}}\!\!\!\left[\spa\prod_{\tiny \tiny\begin{array}{l}i\!\notin\!\!\cD(s)\end{array}}\spa\!\!(1-B_i)\spa\prod_{\tiny \tiny\begin{array}{l}i\!\in\!\!\cD(s)\end{array}}\spa B_i\right]\times\alpha\spa\sum_{\tiny \tiny\begin{array}{l}m\!\in\!\!\cD(s)\end{array}}\spa\sum_{\tiny k=0}^\infty \sum_{n=1}^\infty\frac{ a_n\Gamma\left(k+\frac{(n+1)}{2}\right)/k!}{(1+\frac{\lambda_{_{md}}}{2P\beta\sigma^2})^{k+\frac{(n+1)}{2}}} \left(\frac{c_{_m}}{2}\right)^{k}\clam\no\\\times\left[\frac{\lambda\mdo^{\!\!\spa k+1}}{(\lambda\mdo+\lambda\mdo\frac{c_{_m}}{2})^{k+1}}-\spa\sum_{\tiny \tiny\begin{array}{l}i\!\in\!\!\cD(s)\end{array}}\spa\frac{\lambda\mdo^{\!\!\spa k+1}}{(\lambda\mdo+\lambda\ido
+\lambda\mdo\frac{c_{_m}}{2})^{k+1}}+\spa\spa\sum_{\tiny \tiny\begin{array}{l}i,j\!\in\!\!\cD(s)\\i\neq j\end{array}}\spa\spa \frac{\lambda\mdo^{\!\!\spa k+1}}{~~(\lambda\mdo+\lambda\ido+\lambda\jdo+\lambda\mdo\frac{c_{_m}}{2})^{k+1} }-\ldots\right.
\no\\\left.+(-1)^{|\cD(s)|}\spa\spa\sum_{\tiny \tiny\begin{array}{l}i,j,\ldots,l\!\in\!\!\cD(s)\\i\neq j,\ldots, l\end{array}}\spa\spa \frac{\lambda\mdo^{\!\!\spa k+1}}{~~(\lambda\mdo+\lambda\ido+\lambda\jdo+\ldots \lambda\ldo+\lambda\mdo\frac{c_{_m}}{2})^{k+1}} \right].
\end{eqnarray}
\hrulefill
\setcounter{equation} {29}
\begin{eqnarray}\label{EQ_FINAL_SER_simp}
\bar{P}_e=\frac1{2}B^M+\sum_{l=1}^{M}\!\!\left[\left(\!\!\begin{array}{c}M\\l\end{array}\!\!\right)(1-B)^l B^{M-l}\right]\times\alpha l\times\sum_{\tiny k=0}^\infty
\sum_{n=1}^{n_a}\frac{ a_n\Gamma\left(k+\frac{(n+1)}{2}\right)/k!}{(1+\frac{\lambda}{2P\beta\sigma^2})^{k+\frac{(n+1)}{2}}}
\left(\frac{c}{2}\right)^{k}\left(\frac{\lambda}{2\sigma^2P}\right)^{k+1}\spa\times\sum_{m=1}^l\frac{\left(\!\!\!\begin{array}{c}l-1\\m-1\end{array}\!\!\!\right)}{(m+\frac{c}{2})^{k+1}}.
\end{eqnarray}
\setcounter{equation}{\value{c1}}\hrulefill
\end{figure*}

\subsection{Outage probability conditioned on the decoding set $\cD(s)$}
Conditioned on the decoding set with the old channel realizations, the outage probability with the new CSI is
\begin{eqnarray}\label{EQ_po_DS}
P_o^{\cD(s)}=Pr\left[\hat{\gamma}_{_{md}}\le R_o,\hat{\gamma}_{_{md\!,\!o}}\geq\!\!\!\!\!\!\max_{_{\tiny\begin{array}{l}i\in\!\!\cD(s)\\i\neq m\end{array}}}\!\!\!\{\hat{\gamma}_{_{id\!,\!o}}\}\big|\cD(s)\right].
\end{eqnarray}
Defining
$$
\chi_{_m}\stackrel{\Delta}{=}\max_{\tiny\begin{array}{l}i\in\!\!\cD(s)\\i\neq m\end{array}}\{\hat{\gamma}_{_{id\!,\!o}}\},
$$
and noting that $\chi_{_{m}}$ and $\hat{\gamma}_{_{\!m\!d}}$ are independent, the conditioned outage probability on the decoding set is
\begin{eqnarray}
P_o^{\cD(s)}&=&\int_0^{\infty}Pr\left[\hat{\gamma}_{_{md}}\le R_o\big|\cD(s),\hat{\gamma}_{_{md\!,\!o}}\right]\no\\
&\times&Pr\left[\hat{\gamma}_{_{md\!,\!o}}\geq\chi_{_{m}}\big|\cD(s),\hat{\gamma}_{_{md\!,\!o}}\right]f_{_{\!\hat{\gamma}_{_{md\!,\!o}}}}\spa~(\hat{\gamma}_{_{md\!,\!o}})~d_{_{\hat{\gamma}_{_{md\!,\!o}}}}\no\\
&=&\int_0^{\infty}F_{_{_{\!\!\hat{\gamma}_{_{md}}}}}\spa~(R_o)F_{\!_{\chi_m}}(\hat{\gamma}_{_{md\!,\!o}})f_{_{\!\hat{\gamma}_{_{md\!,\!o}}}}\spa~(\hat{\gamma}_{_{md\!,\!o}})~d_{_{\hat{\gamma}_{_{md\!,\!o}}}}.\label{EQ_INT}
\end{eqnarray}
 Conditioned on $\hat{\gamma}_{_{md\!,\!o}}$ and using (\ref{EQ_hat_est}), $\hat{\gamma}_{_{md}}$ has a non-central Chi square distribution with two degree of freedom and parameter $\eta_{_m}=c_{_m}\hat{\gamma}_{_{md\!,\!o}}$, where $c_{_m}=\frac{2\rho_{_f}^2}{(1-\rho_{_f}^2)\sigma^2_{\hat{h}\mdo}}$b. Therefore we can write $F_{_{_{\!\!\hat{\gamma}_{_{md}}}}}\spa~(x)$ as \cite{Simon1}
\begin{eqnarray}\label{EQ_F_gamma_md}
F_{_{_{\!\!\hat{\gamma}_{_{md}}}}}\spa~(x)\!=\!\sum_{\tiny k=0}^\infty e^{-\!\frac{c_{_m}\lambda\mdo\hat{\gamma}_{_{md\!,\!o}}}{2}}\!\!\left(\!\!\frac{c_{_m}\!\lambda\mdo\hat{\gamma}_{_{md\!,\!o}}}{2}\!\!\right)^{\!\!k}\no\\\times\frac{\gamma(k+1,\frac{\lambda_{_{md}}x}{2\sigma^2})}{(k!)^2},
\end{eqnarray}
where $\gamma(\cdot,\cdot)$ is the \emph{incomplete gamma function} defined as
\begin{equation}
\gamma(s,x)=\int_0^x t^{s-1}e^{-t}dt\no.
\end{equation}
Furthermore, $F_{\!_{\chi_m}}(x)$ is given by
\begin{eqnarray}\label{EQ_F_chi}
F_{\!_{\chi_m}}(x)&=&\spa\spa\prod_{\tiny \tiny\begin{array}{l}i\in\cD(s)\\i\neq m\end{array}}\left[1-e^{(-\lambda\ido x)}\right]\no\\
&=&\!\!\!1-\spa\sum_{\tiny \tiny\begin{array}{l}i\in\cD(s)\\i\neq m\end{array}}\spa e^{-\lambda\ido x}+\spa\spa\sum_{\tiny \tiny\begin{array}{l}i,j\in\cD(s)\\i\neq m,j\end{array}}\spa\spa e^{-(\lambda\ido+\lambda\jdo) x}-\ldots\no\\
&+&(-1)^{|\cD(s)|}\spa\spa\sum_{\tiny \tiny\begin{array}{l}i,j\in\cD(s)\\i\neq m,j,\ldots, l\end{array}}\spa\spa e^{-(\lambda\ido+\lambda\jdo+\ldots \lambda\ldo) x}.
\end{eqnarray}

Using (\ref{EQ_F_gamma_md}) and (\ref{EQ_F_chi}) in (\ref{EQ_INT}) and noting that $f_{_{\!\hat{\gamma}_{_{md\!,\!o}}}}\spa~(x)=\lambda\mdo e^{-\lambda\mdo x}$ and also the fact that\cite{Table-Integral}
$$
\int_0^\infty x^k\exp(-a x)dx=\frac{k!}{a^{k+1}}.
$$
the outage probability conditioned on the decoding set is given by (\ref{EQ_PO_DS_Final}) at the top of the next page.

Finally, substituting (\ref{EQ_DS}) and (\ref{EQ_PO_DS_Final}) into (\ref{EQ_Selection Rule2}), we obtain  (\ref{EQ_OUT_FINAL}) given at the top of the next page.

  For the special case where all the channel estimation errors and delayed feedbacks are the same, i.e., $\lambda=\lambda\smo=\lambda\mdo=\lambda\ido$, and $c_1=c_2=\ldots c_{|\cD(s)|} =c$, (\ref{EQ_OUT_FINAL}) reduces to (\ref{EQ_OUT_FINAL_simp}) given at the top of next page.
\setcounter{equation}{20}
\section{Average Symbol Error Rate}\label{SEC.ASER}
In this section  we derive a closed form expression for the ASER of the DF selection scheme considered in this paper, in the presence of channel estimation error and feedback delay.
Let  the received SNR via the selected path denoted by $\hat{\gamma}_{_{\!m^{\!\star}\!d}}^{\rm eff}=P\hat{\gamma}_{_{\!m^{\!\star}\!d}}$. Then, the ASER can be written as
\begin{equation}\label{EQ_ASER-DF}
\bar{P}_e=\int_0^\infty\alpha Q\left(\sqrt{\beta P\hat{\gamma}_{_{\!m^{\!\star}\!d}}}\right)f_{\hat{\gamma}_{_{\!m^{\!\star}\!d}}}\left(\hat{\gamma}_{_{\!m^{\!\star}\!d}}\right)d\hat{\gamma}_{_{\!m^{\!\star}\!d}}
\end{equation}
where $\alpha$ and $\beta$ depend on the modulation scheme, and $f_{\hat{\gamma}_{_{\!m^{\!\star}\!d}}}$ is the probability density function (PDF) of $\hat{\gamma}_{_{\!m^{\!\star}\!d}}$, which is given by (see Appendix for details)
\begin{eqnarray}\label{EQ_f_BER}
f_{\hat{\gamma}_{_{\!m^{\!\star}\!d}}}(x)&=&\prod_{\tiny i=1}^MB_i\delta(x)\no\\
&+&\spa\spa\sum_{\tiny \tiny\begin{array}{c}\cD(s)\\|\cD(s)|\geq1\end{array}}\!\!\!\left[\prod_{\tiny i\notin\cD(s)}\!\!(1-B_i)\!\!\prod_{\tiny i\in\cD(s)}\!\!B_i\right]f_{\hat{\gamma}_{_{\!m^{\!\star}\!d}|\cD(s)}}(x),\no\\
\end{eqnarray}
where $\delta(x)$ is the delta function and $B_i=\alpha\int_0^{\infty}Q\left(\sqrt{\beta P\hat{\gamma}_{_{sio}}}\right)f_{\hat{\gamma}_{_{sio}}}d\hat{\gamma}_{_{sio}}$ is given as
\begin{equation}\label{EQ_B_I}
B_i=\frac{\alpha}{2}\left[1-\sqrt{\frac{\beta P}{\beta P+2\lambda_{_{sio}}}}\right].
\end{equation}
Here, $f_{\hat{\gamma}_{_{\!m^{\!\star}\!d}|\cD(s)}}(x)$ is the conditional PDF of the received signal via the selected path, which is  given by
\begin{eqnarray}\label{EQ_f_sel_fin}
f_{\hat{\gamma}_{_{\!m^{\!\star}\!d}|\cD(s)}}(x)=\frac{\partial F_{\hat{\gamma}_{_{\!m^{\!\star}\!d}|\cD(s)}}(x)}{\partial x}.
\end{eqnarray}
Noting that
\begin{equation}
F_{\hat{\gamma}_{_{\!m^{\!\star}\!d}|\cD(s)}}(x)=P_o^{\cD(s)}\big|_{R_o=x},\label{EQ_CDF}
\end{equation}
by inserting (\ref{EQ_PO_DS_Final}) into (\ref{EQ_CDF}) and the result in (\ref{EQ_f_sel_fin}) the conditional PDF of the received signal is then given by (\ref{EQ_f_cond}) at the top of the next page. Finding a closed form formula for the integral in (\ref{EQ_ASER-DF}) is not tractable. However,  by substituting (\ref{EQ_f_BER}) into (\ref{EQ_ASER-DF}), we obtain (\ref{EQ_FINAL_SER}) at the top of the next page, where we have used the approximation\cite{Isukapalli2008}
\setcounter{equation}{27}
 \begin{equation}\label{EQ_QAppp}
 Q(x)\approx e^{-\frac{x^2}{2}}\sum_{n=1}^{n_a}a_nx^{n-1},
 \end{equation}
 where
 \begin{equation}\label{A_N}
 a_n=\frac{(-1)^{n+1}(A)^n}{B\sqrt{\pi}(\sqrt{2})^{n+1}n!},
 \end{equation}
  with $A=1.98$ and $B=1.135$.

 In the special case where all the channel estimation errors and delayed feedbacks are the same, i.e., $\lambda=\lambda\smo=\lambda\mdo=\lambda\ido$, and $c_1=c_2=\ldots c_{|\cD(s)|} =c$, (\ref{EQ_FINAL_SER}) reduces to (\ref{EQ_FINAL_SER_simp}) at the top of the page, where $B=B_1=\ldots B_M$.
 \section{Asymptotic Diversity Order}\label{SEC.ADO}
In the previous section, we have derived an exact  ASER expression with feedback delay and channel estimation errors, which is valid for the entire SNR range. To gain further insights into the system's performance, we focus here on the high SNR regime and analyze the asymptotic ASER for the selection scheme under consideration conditioned on the decoding set, i.e., $\bar{P}_e|\cD(s)$\footnote{In the DF scheme, selection relaying exploit the full diversity order, which is equal to the number of relays regardless of the cardinality of the decoding set \cite{Ikki2009,Ikki2010,Beres2008}. Hence, in this work, we focus on the analysis of the asymptotic ASER conditioned on the decoding set}. We assume that all the channel estimation errors and  feedback delay parameters are the same, without loss of generality. Therefore, we have $\lambda=\lambda\smo=\lambda\mdo=\lambda\ido$, and $c_1=c_2=\ldots c_{|\cD(s)|} =c$

Using  (\ref{EQ_f_cond}), we can write the ASER, conditioned on the decoding set, as
\setcounter{equation} {30}
\begin{eqnarray}
\bar{P}_{e|\cD(s)}\spa&=&\spa\alpha\sum_{\tiny k=0}^\infty
\sum_{n=1}^{n_a}\frac{a_n\Gamma\left(k+\frac{(n+1)}{2}\right)/k!}{(1+\frac{\lambda}{2P\beta\sigma^2})^{k+\frac{(n+1)}{2}}}\left(\frac{c}{2}\right)^{\!\! k}
\no\\&\times&\left(\frac{\lambda}{2\sigma^2P\beta}\right)^{k+1}\sum_{m=1}^l\frac{\left(\!\!\!\begin{array}{c}l-1\\m-1\end{array}\!\!\!\right)}{(m+\frac{c}{2})^{k+1}},\no\\
&=&\spa\!\!\sum_{\tiny k=0}^\infty
\sum_{n=1}^{n_a}\frac{g(n,k)}{(1+\frac{\lambda}{2P\sigma^2\beta})^{k+\frac{(n+1)}{2}}}\left(\frac{\lambda}{2\sigma^2P\beta}\right)^{k+1},\no\\\label{EQ_DIV}
\end{eqnarray}
where $g(n,k)$ is a constant depending on $k$ and $n$.

In the following, we consider two different scenarios:
\subsection{Diversity order with  feedback delay and perfect CSI} With perfect CSI, we have $\lambda=1$. Therefore, as $P\rightarrow\infty$, the dominant term in (\ref{EQ_DIV}) would be the term corresponding to $k=0$ and $n=1$, i.e.
\begin{eqnarray}
\bar{P}_{e|\cD(s)}&\propto& \frac1{1+k_0/P}\times\frac1{P},\no\\
&\propto&\frac1{P+k_0},\no\\
&=&O(P^{-1}).
\end{eqnarray}
Thus, in this case, the achievable diversity order in presence of delayed feedback is 1.
\subsection{Diversity order in the presence of feedback delay and imperfect CSI}In the presence of channel estimation error, since $\lambda$ is proportional to $P$, it should be taken into consideration. From (\ref{EQ_DIV}), we have
\begin{eqnarray}
\lim_{p\rightarrow \infty}\bar{P}_{e|\cD(s)}&\propto&\lim_{p\rightarrow \infty}\left[\sum_{\tiny k=0}^\infty
\sum_{n=1}^{n_a}\frac{\left(\frac{\lambda}{P}\right)^{k+1}}{(1+\frac{k_1\lambda}{P})^{k+\frac{(n+1)}{2}}}\right],\no\\
&=&K.
\end{eqnarray}
Hence, the asymptotic diversity order in the presence of  feedback delay and channel estimation errors is 0. This means that the outage probability or the ASER curves are expected to be saturated at high SNR with imperfect CSI.

\section{Average capacity}\label{SEC.CAPAcity}
In this section, we derive an analytical expression for average capacity in the presence of channel estimation error and feedback delay. At the destination and after matched filtering,  the received signal over the selected link is given as
\begin{eqnarray}
\hat{x}&=&\sqrt{P}\hat{h}\msd\spa\!\!^*\quad y\msd,\no\\
       &=&\sqrt{P}\hat{h}\msd\spa\!\!^*\quad (\sqrt{P}h\msd x+n\msd),\no\\
       &=&\sqrt{P}\hat{h}\msd\spa\!\!^*\quad \left[\sqrt{P}\left(\re \hat{h}{\msd}+u\msd\right)x+n\msd\right],\no\\
       &=&\!\!\!\!\!\underbrace{P\re|\hat{h}\msd|^2x}_{\textsf{message~component}}\!\!\!\!+\!\underbrace{P\hat{h}^*{\!\!\!\msd}u\msd x}_{\textsf{error ~component}}+\!\underbrace{\sqrt{P}\hat{h}^*{\!\!\!\msd}n\msd}_{\textsf{noise~component}}. \label{EQ_DR_1}
\end{eqnarray}

Defining
 \begin{subequations}
 \begin{eqnarray}
 &&q\stackrel{\Delta}{=}P\re|\hat{h}\msd|^2,\\
 &&\tilde{n}\stackrel{\Delta}{=}P\hat{h}^*{\!\!\!\msd}u\msd x+\sqrt{P}\hat{h}^*{\!\!\!\msd}n\msd,
 \end{eqnarray}
 \end{subequations}
 (\ref{EQ_DR_1}) can be written as
\begin{equation}
\hat{x}=qx+\tilde{n}.\label{EQ_infothe}
\end{equation}
The capacity conditioned on the decoding set is defined as
 \begin{eqnarray}
C^{\cD(s)}&=&\arg~\max_{f_x(x)}~\cI(\hat{x},x)\no\\
&=&\arg~\max_{f_x(x)}~\{\cH(\hat{x})-\cH(\hat{x}|x)\}\no\\
&=&\arg~\max_{f_x(x)}~\{\cH(\hat{x})-\cH(\tilde{n}|x)\},\label{ED_I_1}
\end{eqnarray}
where $f_x(x)$ is the PDF of the input signal $x$ and $\cH(\cdot)$ is the differential entropy function. It can be deduced easily from (\ref{EQ_infothe}) that $\tilde{n}$ and $x$ are uncorrelated, i.e., ${\rm E}\{\tilde{n}x^*\}=0$. However,  $\tilde{n}$ and $x$ are not independent and therefore $\cH(\tilde{n}|x)\neq\cH(\tilde{n})$. This  means that the standard procedures for deriving  capacity expressions  are not applicable to (\ref{ED_I_1}).
Alternatively, noting that conditioning does not increase the entropy, i.e.,
\[\cH(\tilde{n}|x)\leq\cH(\tilde{n}),\]
we can lower bound the capacity as
\begin{eqnarray}
C^{\cD(s)}\geq\arg~\max_{f_x(x)}~\{\cH(\hat{x})-\cH(\tilde{n})\}.\label{ED_I_2}
\end{eqnarray}
It must be noted that the distribution of $\tilde{n}$ is not Gaussian. However, following \cite{Hassibi}, Theorem 1, we may assume that $\tilde{n}$  has  zero-mean complex Gaussian distribution  with the  same variance, which is the worse case distribution. In this case, we can re-write (\ref{ED_I_2})  as
\begin{eqnarray}
C^{\cD(s)}\geq\arg~\max_{f_x(x)}~\{\cH(\hat{x})-\cH(\tilde{n})\}\geq\arg~\max_{f_x(x)}~\{\no\\\cH(\hat{x})-\cH({n}^\dagger)\},\label{EQ_inofi}
\end{eqnarray}
where $n^{\dagger}\sim\mathcal{CN}\left(0,P^2|\hat{h}\msd|^2\sigma^2\msd+P|\hat{h}\msd|^2\right)$. Maximizing (\ref{EQ_inofi}) with respect to $f_x(x)$ and using (\ref{EQ_infothe}),  a lower bound on the average capacity, conditioned on the decoding set, can be written as
\begin{eqnarray}
C^{\cD(s)}&\geq& C^{\cD(s)}_{\rm lb}\no\\&&=\frac1{2}\log\left(1+\frac{q^2}{P^2|\hat{h}\msd|^2\sigma^2\msd+P|\hat{h}\msd|^2}\right),\no\\
&&=\frac1{2}\log\left(1+P\hat{\gamma}\msd\right),\label{EQ_cap_lb}
\end{eqnarray}
where $\hat{\gamma}\msd=\frac{\re^2|\hat{h}\msd|^2}{1+P\sigma^2_{u_{\!m^{\!\star}\! d}}}$. Hence, a lower bound on the   S-DF capacity  can be obtained  as
\begin{eqnarray}
C_{\rm lb}=\sum_{\cD(s)}C_{\rm lb}^{\cD(s)}Pr[\cD(s)].
\end{eqnarray}
Therefore, the average   capacity bound, i.e., $\bar{C}_{\rm lb}$, can be written as
\begin{eqnarray}\label{EQ_capacity_lower}
&&\bar{C}_{\rm lb}=\frac1{2}\sum_{\cD(s)}Pr[\cD(s)]\no\\&&\times\int_0^{\infty}\log\left(1+P\hat{\gamma}\msd\right)f_{\hat{\gamma}_{_{\!m^{\!\star}\!d}|\cD(s)}}(\hat{\gamma}\msd)d\hat{\gamma}\msd.
\end{eqnarray}
Substituting (\ref{EQ_f_cond}) into (\ref{EQ_capacity_lower}) yields (\ref{EQ_average_capacity}) given at the top of the next page.
 Note that in our derivation we have used the integration formula \cite{Table-Integral}
\begin{eqnarray}
&&\!\!\!\int_0^{\infty}\ln(1+ax)x^k e^{-x}dx=\sum_{\mu=0}^k \frac{k!}{(k-\mu)!}\no\\
&&\spa\spa\!\!\!\times\left[\frac{(-1)^{k-\mu-1}e^{1/a}}{a^{k-\mu}}\rE i\left(-\frac1{a}\right)+\sum_{t=1}^{k-\mu}(t-1)!\left(-\frac1{a}\right)^{k-\mu-t}\right],\no \\
\end{eqnarray}
where
\[\rE i(\gamma)\stackrel{\Delta}{=}\int_\gamma^{\infty}\frac1{x}\exp(-x)dx\]
is the \emph{exponential integral function}.

In the special case where all the channel errors and feedback delay parameters are the same, the average capacity bound can be obtained as in (\ref{EQ_cap_simp}) at the top of next page.

\begin{figure*}
\begin{eqnarray}\label{EQ_average_capacity}
\bar{C}_{\rm lb}\!=\frac1{\ln2}\!\!\sum_{\tiny \tiny\begin{array}{c}\cD(s)\end{array}}\spa\left[\!\!\!\prod_{\tiny \tiny\begin{array}{c}\!m\!\!\in\!\!\!\cD(s)\end{array}}\spa\exp\left(-\lambda\smo R_o\right)
\spa\prod_{\tiny \tiny\begin{array}{c}\!\!m\!\!\notin\!\!\!\cD(s)\end{array}}\spa\left[1-\exp\left(-\lambda\smo R_o\right)\right]\right]\times\spa\sum_{\tiny \tiny\begin{array}{l}m\!\!\in\!\!\cD(s)\end{array}}\!\!\!\sum_{k=0}^\infty \sum_{\mu=0}^k \frac{\left(\frac{c_m}{2}\right)^k}{(k-\mu)!}\left[\frac{(-1)^{k-\mu-1}}{\left(\frac{2\sigma^2P}{\lambda_{_{\!md}}}\right)^{k-\mu}}e^{\left(\frac{\lambda_{_{\!md}}}{2\sigma^2P}
\right)}\right.\no\\
\left.\times\rE i\left(-\frac{\lambda_{_{\!md}}}{2\sigma^2P}\right)+\sum_{t=1}^{k-\mu}(t-1)!\left(-\frac{\lambda_{_{\!md}}}{2\sigma^2P}\right)^{k-\mu-t}\right]\times
\left[\frac{\lambda\mdo^{\!\!\spa k+1}}{(\lambda\mdo+\lambda\mdo\frac{c_{_m}}{2})^{k+1}}-\spa\sum_{\tiny \tiny\begin{array}{l}i\!\in\!\!\cD(s)\end{array}}\spa\frac{\lambda\mdo^{\!\!\spa k+1}}{(\lambda\mdo+\lambda\ido
+\lambda\mdo\frac{c_{_m}}{2})^{k+1}}\right.\no\\
+\left.\spa\spa\sum_{\tiny \tiny\begin{array}{l}i,j\!\in\!\!\cD(s)\\i\neq j\end{array}}\spa\spa \frac{\lambda\mdo^{\!\!\spa k+1}}{~~(\lambda\mdo+\lambda\ido+\lambda\jdo+\lambda\mdo\frac{c_{_m}}{2})^{k+1} }-\ldots\right.
\left.+(-1)^{|\cD(s)|}\spa\spa\sum_{\tiny \tiny\begin{array}{l}i,j,\ldots,l\!\in\!\!\cD(s)\\i\neq j,\ldots, l\end{array}}\spa\spa \frac{\lambda\mdo^{\!\!\spa k+1}}{~~(\lambda\mdo+\lambda\ido+\lambda\jdo+\ldots \lambda\ldo+\lambda\mdo\frac{c_{_m}}{2})^{k+1}} \right].
\end{eqnarray}
\hrulefill
\begin{eqnarray}\label{EQ_cap_simp}
\bar{C}_{\rm lb}=\frac1{\ln2}\sum_{l=1}^M\exp\left(-l\lambda R_o\right)\left[1-\exp\left(-\lambda R_o\right)\right]^{M-l}\times l\times\sum_{\tiny k=0}^\infty \sum_{\mu=0}^k \frac{\left(\frac{c}{2}\right)^k}{\lambda^{k+1}(k-\mu)!}\times\left[\frac{(-1)^{k-\mu-1}}{\left(\frac{2\sigma^2P}{\lambda}\right)^{k-\mu}}e^{\left(\frac{\lambda}{2\sigma^2P}\right)}\rE i\left(-\frac{\lambda}{2\sigma^2P}\right)\right.\no\\\left.+\sum_{t=1}^{k-\mu}(t-1)!\left(-\frac{\lambda}{2\sigma^2P}\right)^{k-\mu-t}\right] \left(\frac{c}{2}\right)^{k}\times\sum_{m=1}^l\frac{\left(\!\!\!\begin{array}{c}l-1\\m-1\end{array}\!\!\!\right)}{(m+\frac{c}{2})^{k+1}}.
\end{eqnarray}
\hrulefill
\end{figure*}

\section{Simulation Results}\label{SEC.Simulation R}

In the section, we investigate the performance of selection cooperation in the presence of channel estimation errors and feedback delay through  Monte-Carlo simulation.
The transmitted symbols are drawn from an antipodal BPSK constellation, which means, $\alpha=1$ and $\beta=2$.
The node-to-node channels are assumed to be zero mean independent Gaussian processes, with variance $\sigma^2_h=\sigma^2_{h_o}=1-\sigma^2_e$. In this case, $\hat{h}$ and $\hat{h}_o$ are zero mean Gaussian processes with unit variance and $\rho_e=1-\sigma^2_e$. The variance of noise components is set to $N_0=1$ and $R_o=1$ bps/Hz.

\begin{figure}[htb!]
   \begin{center}
    \begin{tabular}{c}
       \psfig{figure=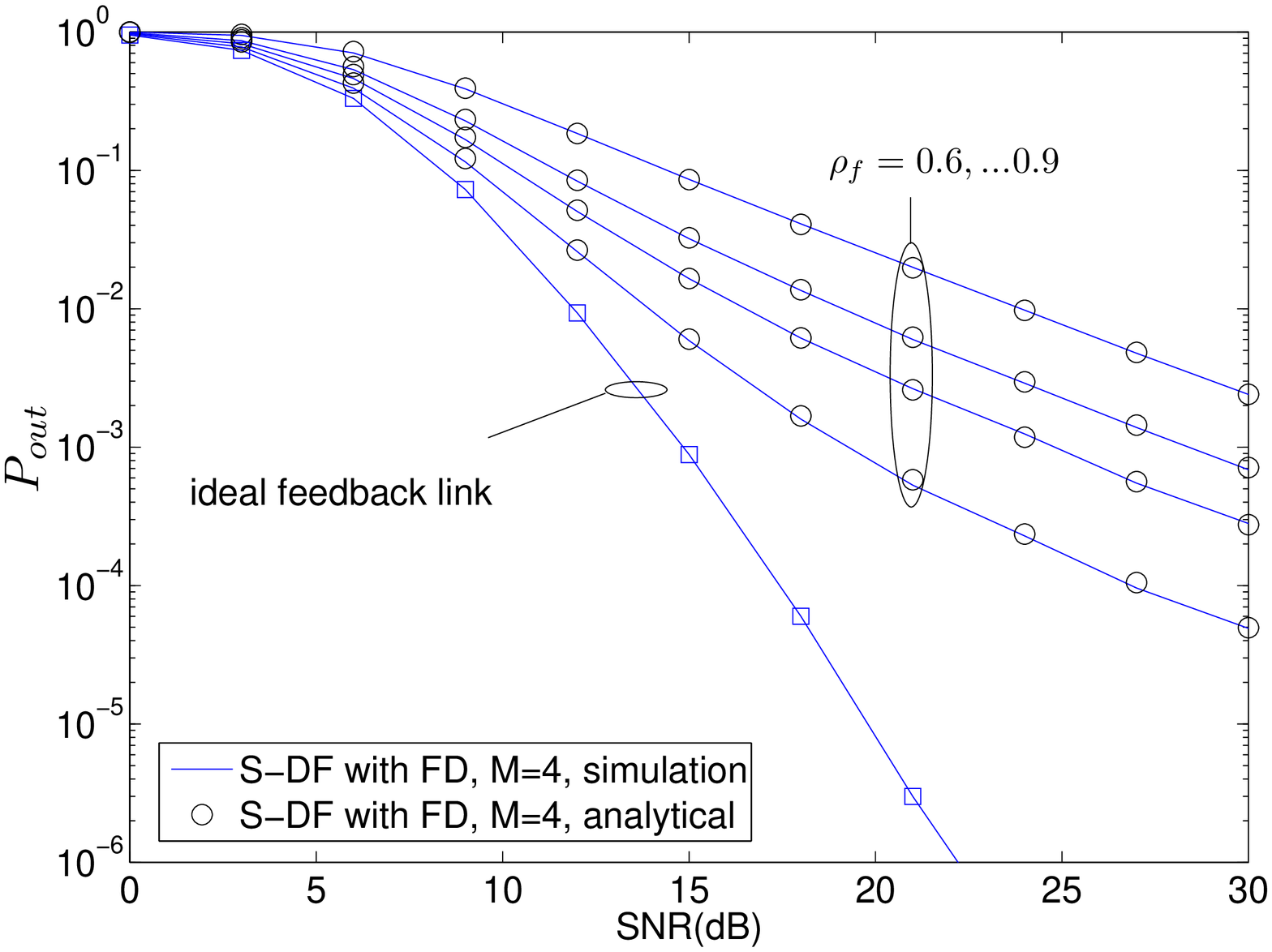,width=5.2in,height=2.82in} \\
    \end{tabular}
   \end{center}
   \vskip -11pt
 \caption{\!\!\!\!\!Outage Probability for $M=4$ and perfect CSI in presence of delay in the feedback link. In this figure $P_{_e}=1$ and $\rho_{_f}=0.6, 0.7, \ldots 1$.}
\label{Fig_M1} \vskip -11pt
\end{figure}

\emph{Outage Probability}:
Fig.~\ref{Fig_M1} shows the performance of the system with $M=4$ relays with  feedback delay (FD) for $\rho_{_f}=0.6,0.7,0.8,0.9$ and 1.
In Fig.~\ref{Fig_M1}, we assume  perfect CSI, i.e., $\rho_e=1$. We observe a perfect match between the analytical and  simulation results. It can also be deduced from the slope of the curves  that, for $M=4$ and  $\rho_{_f} < 1$, the diversity order of the selection scheme for is 1. In the case of ideal feedback link, i.e., $\rho_{_f}=1$, the diversity order of 4 is observed.

\begin{figure}[htb!]
   \begin{center}
    \begin{tabular}{c}
       \psfig{figure=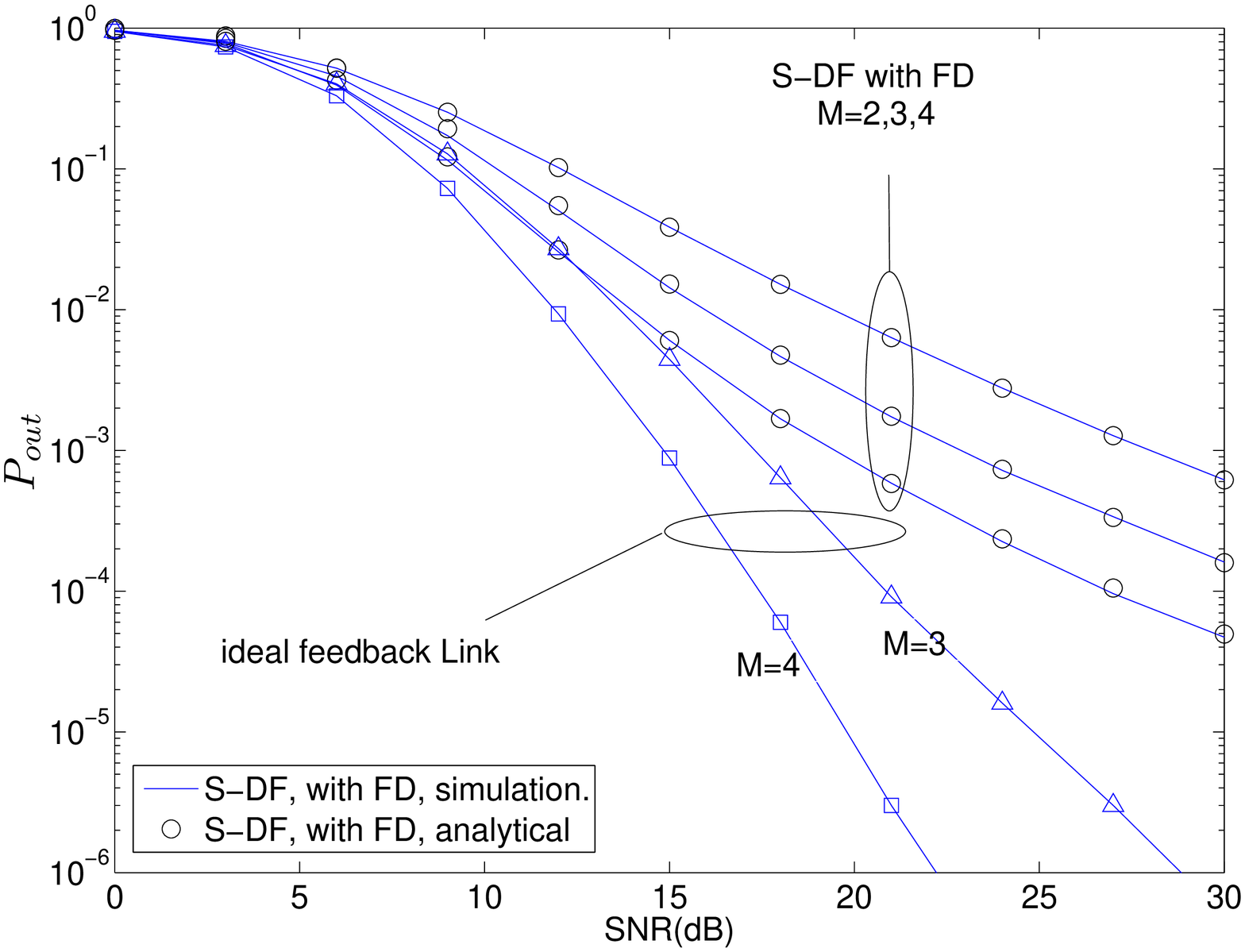,width=5.2in,height=2.82in} \\
    \end{tabular}
   \end{center}
   \vskip -11pt
 \caption{\!\!\!\!\!Outage Probability for $M=2,3,4$ and perfect CSI in presence of delay in the feedback link. In this figure $\rho_{_e}=1$ and $\rho_{_f}=0.9$.}
\label{Fig_M2} \vskip -11pt
\end{figure}

Fig.~\ref{Fig_M2} illustrates the performance of the system for $M=2, 3, 4$, with $\rho_{_f} = 0.9, 1$. In Fig.~\ref{Fig_M2}, the same slope for different number of relays is noticed, confirming our analytical observations. For the case of $\rho_{_f}=1$ and choosing $M=3, 4$, the diversity orders of 3 and 4 are observed, respectively.

\begin{figure}[htb!]
   \begin{center}
    \begin{tabular}{c}
       \psfig{figure=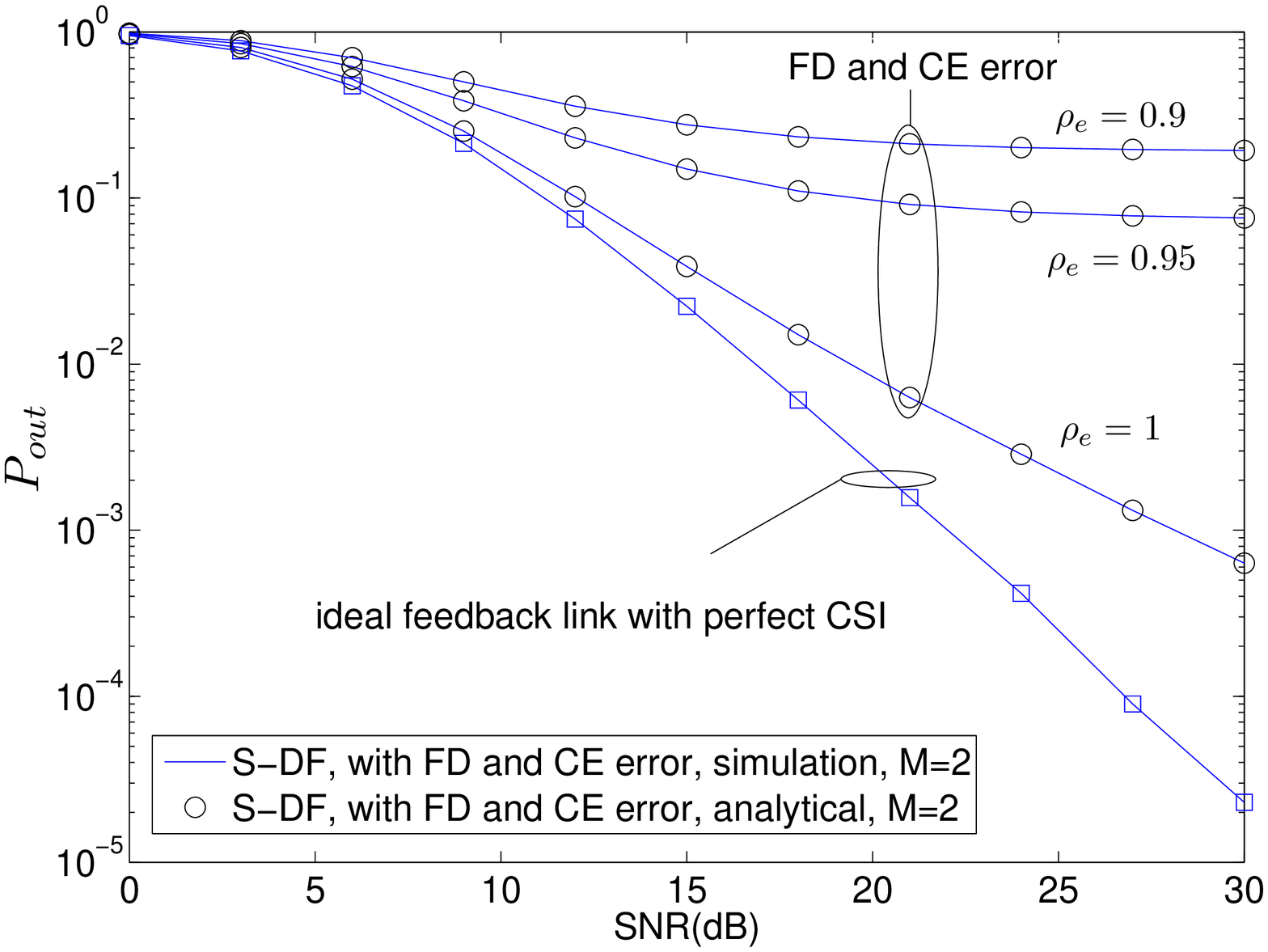,width=5.2in,height=2.82in} \\
    \end{tabular}
   \end{center}
   \vskip -11pt
 \caption{\!\!\!\!\!Outage probability for $M=2$ presence of delay in the feedback link and channel estimation error. In this figure $\rho_{_f}=0.9$.}
\label{Fig_M3} \vskip -11pt
\end{figure}

In Fig.~\ref{Fig_M3}, we study the effect of  both feedback delay and channel estimation errors. An error floor, in presence of channel estimation error, is noticed, as predicted by (36).

\begin{figure}[htb!]
   \begin{center}
    \begin{tabular}{c}
       \psfig{figure=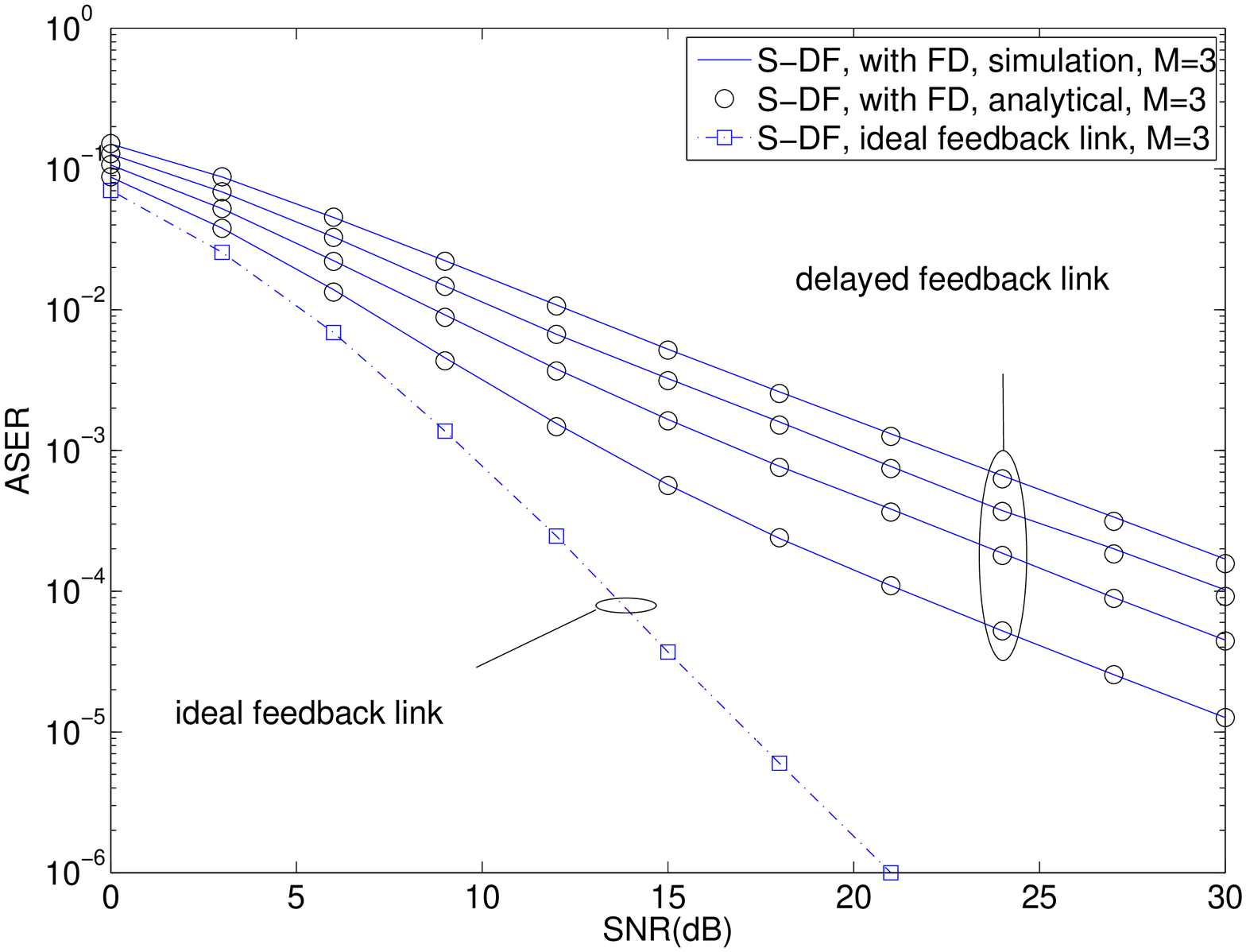,width=5.2in,height=2.82in} \\
    \end{tabular}
   \end{center}
   \vskip -11pt
 \caption{\!\!\!\!\!ASER for $M=3$ and perfect CSI in presence of delay in the feedback link. In this figure $\rho_{_e}=1$ and $\rho_{f}=0.6,0.7,\ldots,1$.}
\label{Fig_M4} \vskip -11pt
\end{figure}

\emph{ASER performance}: Fig.~\ref{Fig_M4} shows the ASER  in the presence of  feedback delay for $M=3$. We assume  perfect CSI knowledge.  Assuming an ideal feedback link, the full diversity order of  3  is achieved. However, with feedback delay, the diversity order is reduced to 1, as predicted earlier.

\begin{figure}[htb!]
   \begin{center}
    \begin{tabular}{c}
       \psfig{figure=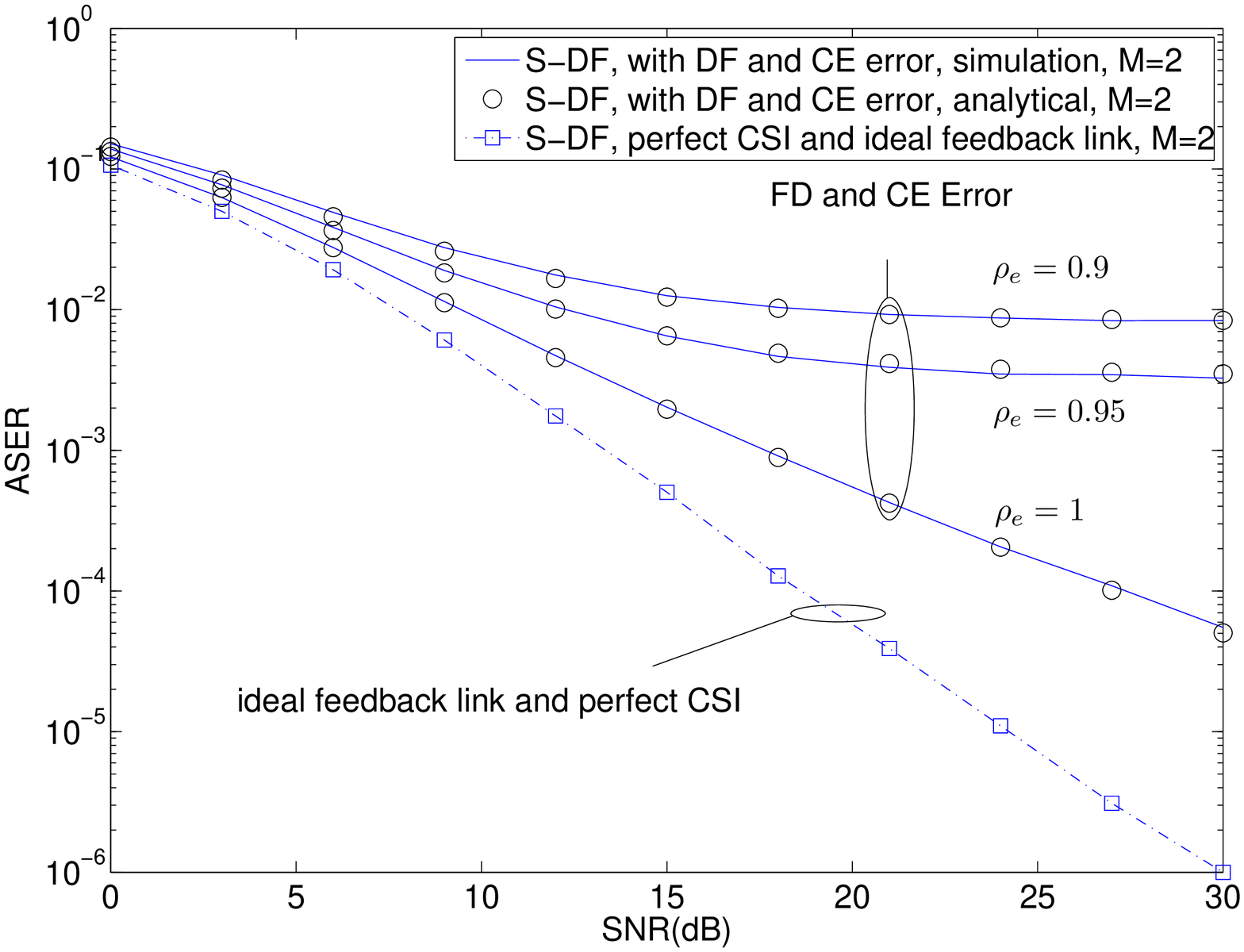,width=5.2in,height=2.82in} \\
    \end{tabular}
   \end{center}
   \vskip -11pt
 \caption{\!\!\!\!\!ASER for $M=2$ presence of delay in the feedback link and channel estimation error. In this figure $\rho_{_f}=0.9$.}
\label{Fig_M5} \vskip -11pt
\end{figure}

Fig.~\ref{Fig_M5} illustrates the performance of the ASER in presence of channel estimation error and feedback delay for $M=2$. It is clear from Fig.~\ref{Fig_M5} that channel estimation errors reduce the diversity order of the system to zero, confirming our earlier analysis.

\emph{Diversity order}:  Noting that the asymptotical diversity order $d$ is given by the magnitude of the
slope of ASER against average SNR in a log-log scale \cite{Nabar2004}:
\begin{eqnarray}
d=\lim_{{\tt SNR}\rightarrow \infty}-\frac{\log\left(\bar{P}_{e}\right)}{\log{\tt SNR}}.\label{EQ_outdiv}
\end{eqnarray}
in this subsection we analyze the asymptotical diversity order of the underlying selection scheme. We  assume perfect CSI knowledge, unless otherwise indicated.

\begin{figure}[htb!]
   \begin{center}
    \begin{tabular}{c}
       \psfig{figure=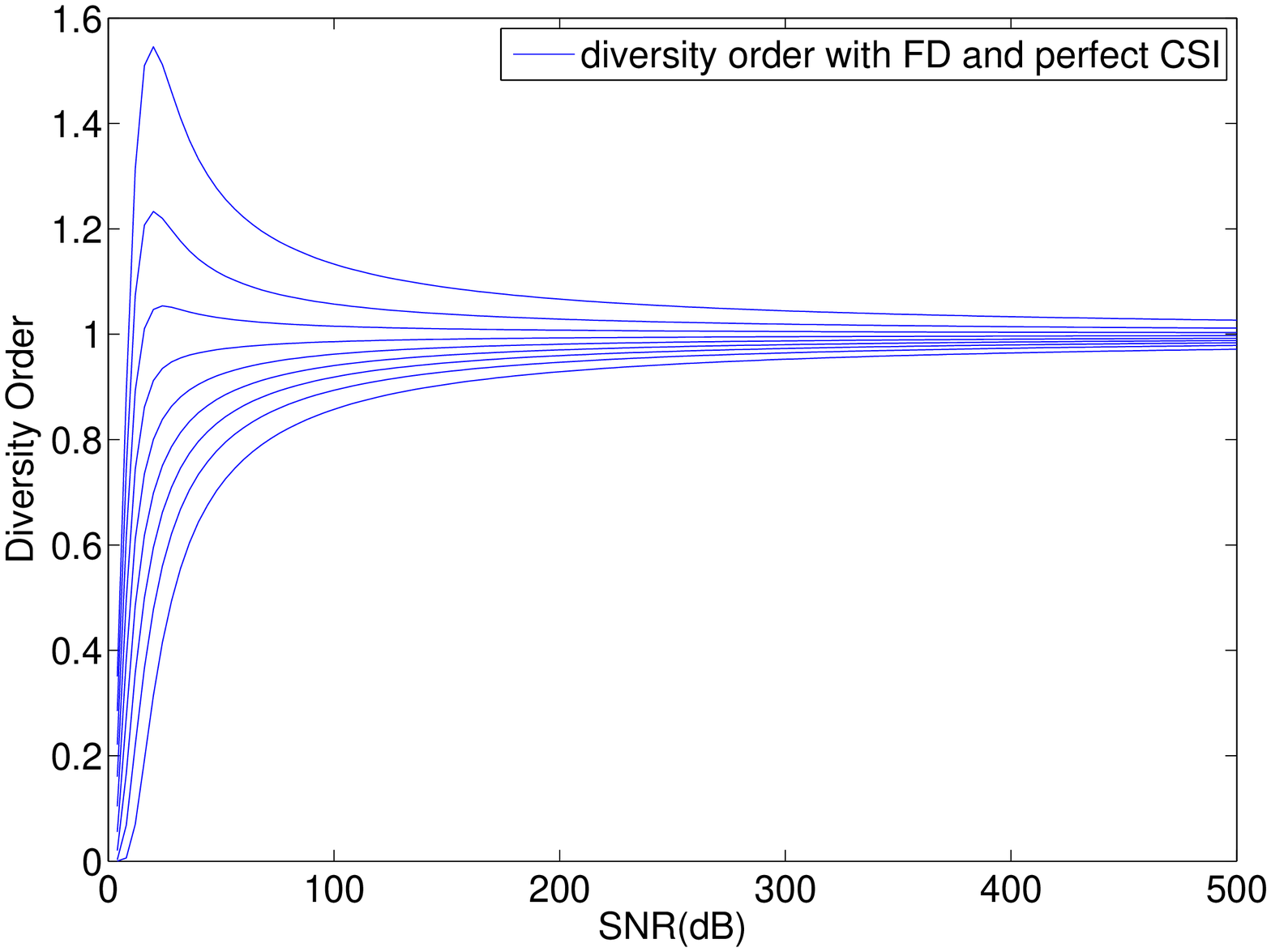,width=5.2in,height=2.82in} \\
    \end{tabular}
   \end{center}
   \vskip -11pt
 \caption{\!\!\!\!\!Effective diversity order versus SNR. $\rho_{_e}=1$ and $M=4$, The figure is plotted for different values of $\rho_{f}$ and is obtained from (\ref{EQ_outdiv}).}
\label{Fig_M6} \vskip -11pt
\end{figure}

Fig.~\ref{Fig_M6} shows the diversity performance of the system  in presence of feedback delay link for different values of $\rho_{_f}$. We  assume perfect CSI knowledge. It is observed that the asymptotical diversity order of the system tends to 1 for all $\rho_f$ values. This illustrates the destructive effect of feedback delay  and demonstrates that relay selection  in this case is annihilated.

\begin{figure}[htb!]
   \begin{center}
    \begin{tabular}{c}
       \psfig{figure=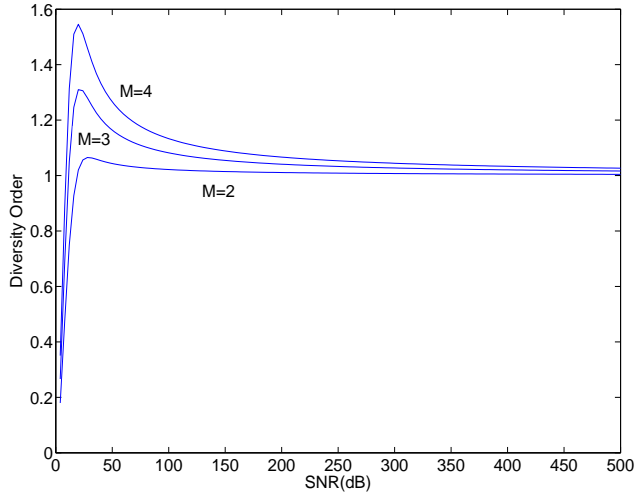,width=5.2in,height=2.82in} \\
    \end{tabular}
   \end{center}
   \vskip -11pt
 \caption{\!\!\!\!\!Effective diversity order versus SNR. $\rho_{_e}=1$, $\rho_{_f}=0.9$ and $M=2,3,4$. The figure is obtained from (\ref{EQ_outdiv}).}
\label{Fig_M7} \vskip -11pt
\end{figure}

Fig.~\ref{Fig_M7} illustrates the asymptotical diversity  for different number of relays i.e., $M=2,3,4.$ It is obvious that at high SNR the diversity order is independent of the number of relays in the system.

\begin{figure}[htb!]
   \begin{center}
    \begin{tabular}{c}
       \psfig{figure=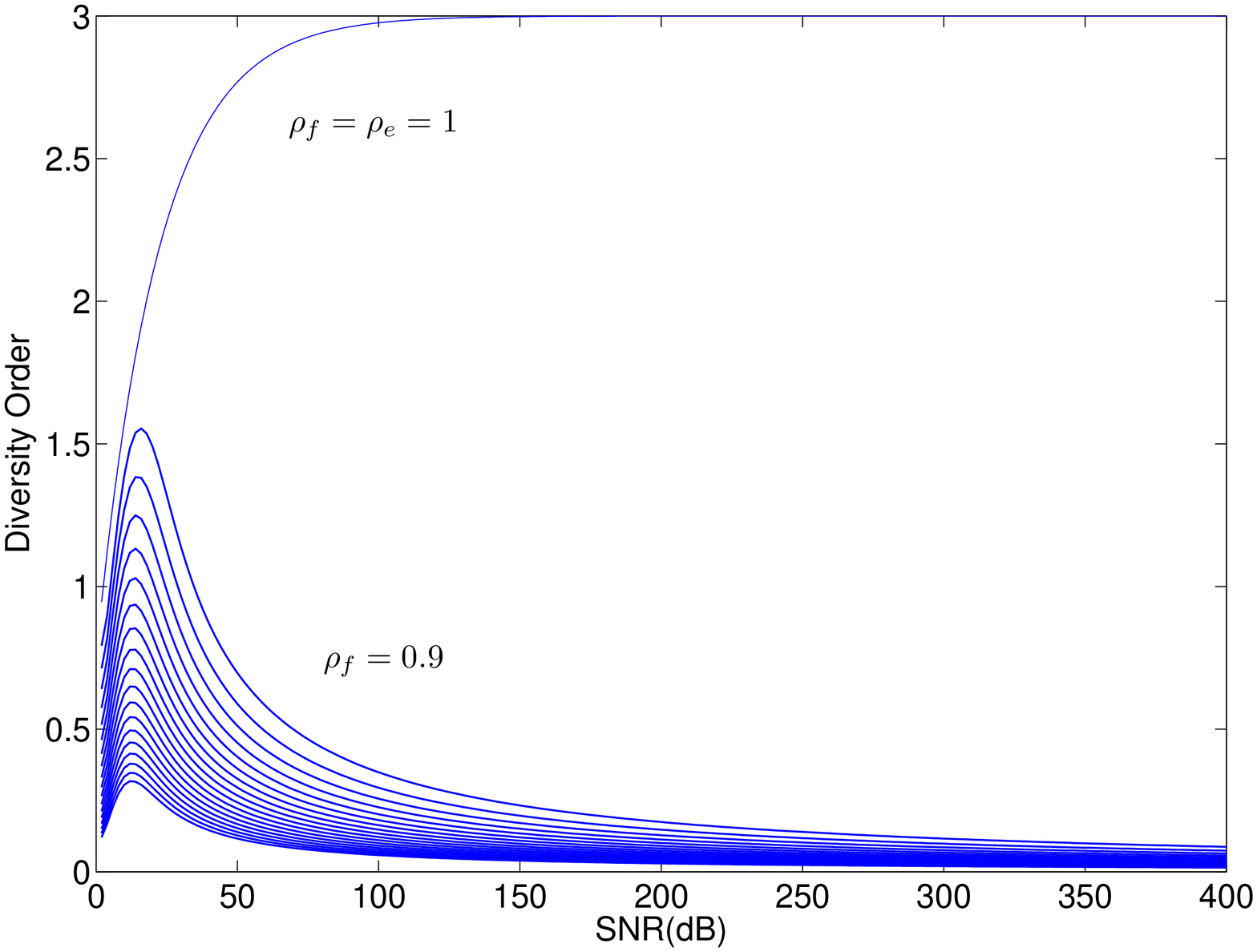,width=5.2in,height=2.82in} \\
    \end{tabular}
   \end{center}
   \vskip -11pt
 \caption{\!\!\!\!\!Effective diversity order versus SNR. $\rho_{_f}=0.9$ and $M=3$, The figure is plotted for different values of $\rho_{e}$ and is obtained from (\ref{EQ_outdiv}).}
\label{Fig_M8} \vskip -11pt
\end{figure}

Fig.~\ref{Fig_M8} depicts the asymptotical diversity order in presence of channel estimation errors. It is noticed that the asymptotical  diversity order in this case is reduced to zero, confirming our earlier observations in Figs.~\ref{Fig_M3} and \ref{Fig_M5}.

\begin{figure}[htb!]
   \begin{center}
    \begin{tabular}{c}
       \psfig{figure=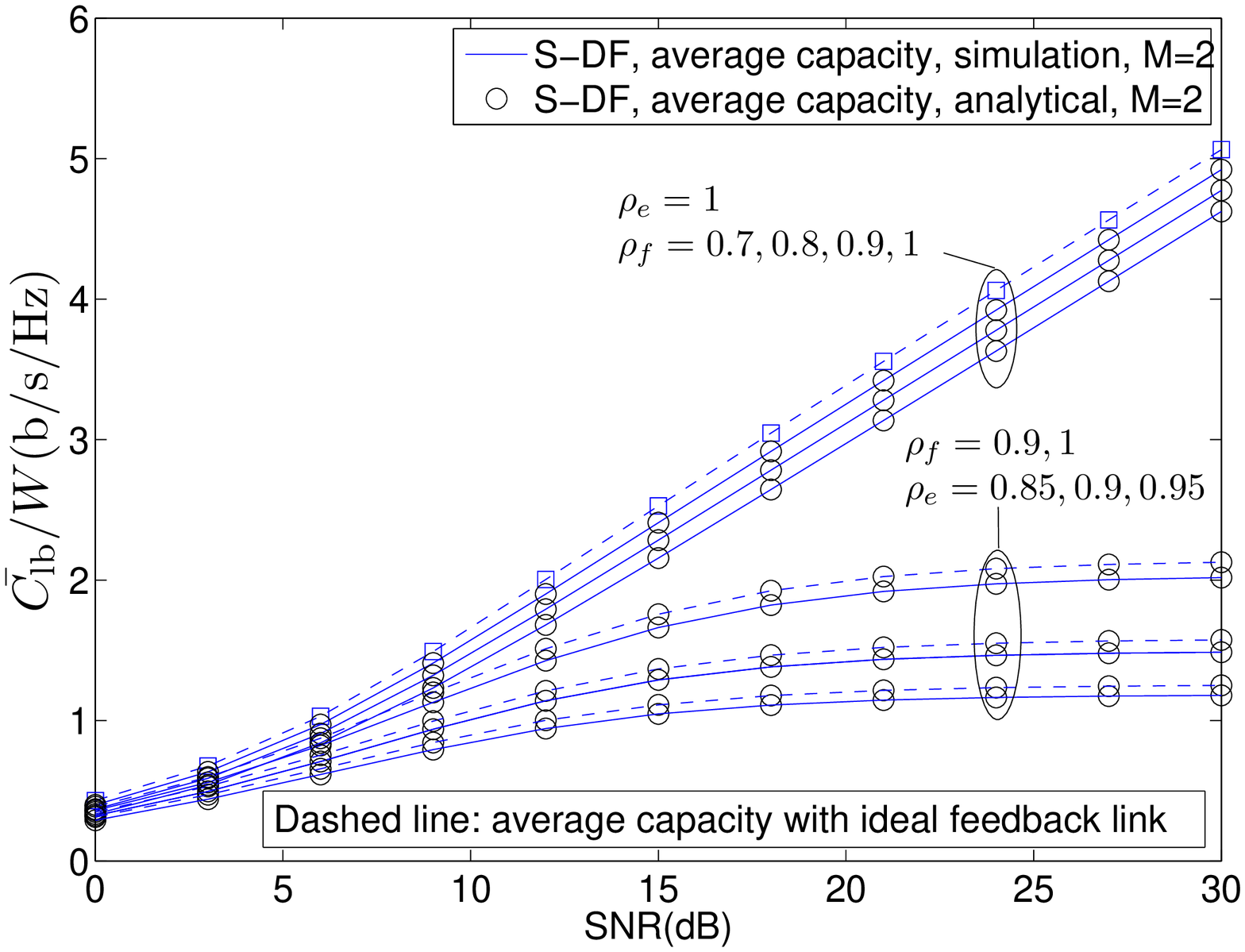,width=5.2in,height=2.82in} \\
    \end{tabular}
   \end{center}
   \vskip -11pt
 \caption{\!\!\!\!\!Lower bound average capacity versus SNR for different values of $\rho_f$ and $\rho_e$.}
\label{Fig_M9} \vskip -11pt
\end{figure}

\emph{Average Capacity}:~Fig.~\ref{Fig_M9} shows the lower bound average capacity in bits per second per Hz per bandwidth versus SNR. It is observed that the presence of channel estimation errors result in  capacity ceilings in the average capacity curves. It can be also seen that feedback delay  aggravates the average capacity performance of the system. However, channel estimation errors have a greater effect  in worsening the average capacity performance of the system.

\hspace{82mm}$\blacksquare$



\setcounter{section}{10}
\section{Conclusion}\label{SEC.CoNC}
In this paper, we discuss a relay selection scheme in DF networks. We show that the presence of  channel estimation errors and  feedback delay degrades the performance and also reduces the  diversity order  of S-DF.
We derive an exact analytical expressions for the outage probability, average symbol error rate and  average capacity bound.

\section*{Appendix\\Deriving the Probability density function of $\gamma\msd$}
Let $\gamma\msd$ denote the normalized received SNR at the destination terminal over the  $S\rightarrow R_{i^{\!\star}}\rightarrow D$ link. Then, the PDF of $\gamma\msd$ is written as
\begin{eqnarray}\label{EQ_APPENDIX_PDF}
f_{\gamma\msd}(x)=Pr(\text{\footnotesize all relays are off})f_{\gamma\msd{\big|(\text{all relays are off})}}(x)\no\\+\spa\!\!\!\sum_{\tiny \tiny\begin{array}{c}\cD(s)\\|\cD(s)|\geq1\end{array}}\spa \!\!\!Pr(\text{\footnotesize relays in}~\cD(s)~\text{\footnotesize are on})f_{\gamma\msd\big|(\text{relays in}~\cD(s)\text{are on})}(x).
\end{eqnarray}
The $i^{th}$ relay decodes its received signal erroneously with probability $B_i$ which is given in (\ref{EQ_B_I}). Since all $S\rightarrow R_i$ links are statistically, we have
\begin{equation}\label{EQ_BBI}
Pr(\text{\footnotesize all relays are off})=\prod_{i=1}^MB_i.
\end{equation}
On the other hand, if all the relays are off, then no communication would occur between source and destination terminal.  The received SNR  at the destination terminal would be zero. Therefore, the conditional PDF can be written as \cite{Beauliu2006}
\begin{equation}\label{EQ_Condd}
f_{\gamma\msd{\big|(\text{all relays are off})}}(x)=\delta(x).
\end{equation}

The probability of decoding set, given that there is at least one relay in $\cD(s)$, is given by
\begin{equation}\label{EQ_APPE(DS)}
Pr(\text{\footnotesize relays in}~\cD(s)~\text{\footnotesize are on})=\left[\prod_{\tiny i\notin\cD(s)}\!\!(1-B_i)\!\!\prod_{\tiny i\in\cD(s)}\!\!B_i\right].
\end{equation}
Inserting (\ref{EQ_BBI}), (\ref{EQ_Condd}) and (\ref{EQ_APPE(DS)}) in (\ref{EQ_APPENDIX_PDF}) yields (\ref{EQ_f_BER}).

\end{document}